\documentclass{article}

\usepackage{arxiv}

\usepackage[utf8]{inputenc} 
\usepackage[T1]{fontenc}    
\usepackage[hidelinks]{hyperref}      
\usepackage{url}            
\usepackage{booktabs}       
\usepackage{amsfonts}       
\usepackage{amsmath}
\usepackage{nicefrac}       
\usepackage{microtype}      
\usepackage{lipsum}		    
\usepackage{graphicx}
\usepackage{natbib}
\usepackage{doi}
\usepackage{rotating}
\usepackage{makecell}
\usepackage{xcolor}
\usepackage[normalem]{ulem}
\usepackage{soul}
\usepackage{longtable}
\usepackage{array} 
\usepackage{booktabs}
\usepackage{enumitem} 
\usepackage{comment}
\usepackage{caption}
\usepackage{adjustbox}
\usepackage{floatrow}

\graphicspath{ {./Figures/} }

\usepackage{floatrow}
\usepackage{caption}
\usepackage{subcaption}
\usepackage[rightcaption]{sidecap}
\usepackage{longtable,booktabs,array}

\title{A Stock-Flow Framework for Editorial Board Dynamics: The Case of Economics Journals, 1866–2019 
}

\date{} 					

\author{\href{https://orcid.org/0000-0003-0293-482X}{\includegraphics[scale=0.06]{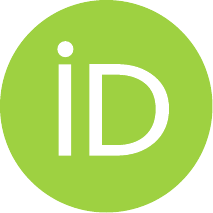}\hspace{1mm}Alberto Baccini}
  \\
	Dipartimento di Economia Politica e Statistica\\
	Università degli Studi di Siena\\
	Siena, Italy \\
	\texttt{alberto.baccini@unisi.it} \\
	}

\renewcommand{\headeright}{}
\renewcommand{\undertitle}{}
\renewcommand{\shorttitle}{A Stock-Flow Framework for Editorial Board Dynamics}

\hypersetup{
pdftitle={},
}

\makeatletter
\input{size12.clo}
\makeatother

\begin{document}

\maketitle

\begin{abstract}
Research on the editorial boards of scholarly journals has predominantly relied on static, cross-sectional data, focusing on their composition or interlocking editorships at single points in time. To address this gap, a formal stock-flow framework is developed for analyzing the longitudinal dynamics of editorial boards. The model integrates three interconnected layers: journal demographics, the dynamics of editorial positions, and the dynamics of board members. This framework is applied to the Gatekeepers of Economics Longitudinal Database (GOELD), which contains annual snapshots of editorial boards for approximately 1,700 economics journals from 1866 to 2006 (by decade), plus the years 2012 and 2019. The period until 1946 was characterized by small-scale: few journals and compact editorial communities. The decade from 1946 to 1956 marked the shift toward a ``big science'' model, initiating an era of expansionary growth fueled primarily by the founding of new journals. The contemporary period (2006-2019) appears to represent a structural break, characterized by low flux and more stable and more closed editorial communities. The results shows that the proposed framework enables a dynamic, long-term analysis of how journals and their gatekeeping systems evolve, grow, and structure themselves.

\end{abstract}

. 

\newpage

\section{Introduction}

Scholarly journals have become the primary instrument for disseminating research, yet paradoxically, little is known about their internal governance and organizational structures. Significant attention has been paid to peer-review processes, particularly to their qualitative aspects and effectiveness. However, editorial hierarchies, decision-making roles, and operational mechanisms that govern journals still constitute a major gap in our understanding of contemporary science communications. 

Building on this focus, the editorial board emerges as the primary unit for analyzing the  internal governance of journals. Following the seminal work of \citet{crane1967}, boards are recognized as the core gatekeepers of knowledge, wielding decisive power over which articles enter the scholarly record. Consequently, much of the literature has focused on the relationship between board composition and the outcomes of their decisions. This includes analyzing potential biases in manuscript selection, for instance by examining the publication rates of papers authored by editors themselves or by scholars affiliated with the institutions of board members. In an evaluative turn, scholars like \citet{braun2005a} have even attempted to use editorial board composition as a proxy for journal impact, investigating its correlation with citation metrics. A more recent approach analyzes the interlocking editorship, that is the network structure created by editors serving on multiple boards simultaneously \citep{baccini2009}. This framework allows for the mapping of scientific communities, ``invisible colleges'' and academic elites, identifying the structural characteristics of the network of relations linking journals and scholars in their gatekeeping activities. 

This paper is focused on editorial boards of journals. Its core intuition is that scientific journals can be studied as organizations: the analysis of the evolution over time in the demographics of journals and of their editorial boards could provide insights into publishing landscape of a scientific field and internal governance of journals. The specific focus adopted here is on analyzing the journal demographics and the dynamics of editorial boards through a stock-flow framework. This intuition is inspired by the analysis of firm demographics and labour market, which studies the birth and death of firms and the resulting processes of job creation and job destruction \citep[for all, see the classic][]{jobcreation}. Analogously, the demographic dynamics of journals can be analyzed in reference to the creation and destruction of \textit{seats} in their editorial board. Indeed, the creation of new seats on boards can result from the internal expansion of an existing journal or from the founding of new journals. Conversely, the destruction of seats can stem from the contraction of existing boards or the discontinuation of journals. 
Differently from the analysis of job creation and destruction, our analysis can also track the dynamics of the editors themselves, that is the flows of members entering or leaving boards. These flows represent the dynamic basis for the phenomena of interlocking editorship.

This paper has two main objectives. The first is to systematize the study of journal demographics and board dynamics within a coherent stock-flow analytical framework. This framework identifies the key indicators for the empirical analysis of seat creation and destruction flows, as well as editor entry and exit. The second is to test the application of this analytical framework to data relating to the editorial boards of economics journals over the long term, from the appearance of the first recognized journals (1866) to the present day (2019). 

The primary research question guiding this study is whether the analytical technique proposed here can reveal significant and hitherto underexplored dynamics within the world of scientific journals. The successful answer to this question would provide a quantitative framework for further analysis of the role of journals as the main institutions responsible for the gatekeeping of science, thereby aiding in the interpretation of how the organisation of science evolves.

The paper is organized as follows. After a summary reconstruction of the related literature in Section 2, Section 3 develops the stock-flow models of journals demographics, of seat creation and destruction, and of membership. Section 4 presents the data used for the empirical application of the stock-flow analysis. The results of the empirical analysis are presented in Section 5. Section 6 discusses these results, and Section 7 briefly concludes by suggesting further possible research perspectives.

\section{Related works}

The theme of this paper is at the cross-road between two different literature. The first one regards specifically the demographics of journals in science, and the second their internal structure, that is their editorial boards.

As for the demographics of journals, the available analysis are mainly concentrated on the trend of their number over time, at least since the famous graph in which \citet{price} represented the number of scientific journals from 1600 to 1960. In a recent study, Mabe (2003) documented a remarkably constant long-term growth rate of approximately 3.5\% per annum in the number of active scientific journals, as recorded in Ulrich's Periodicals Directory. Analysis of the 20th century revealed three distinct phases: a stable pre-1940 growth (3.23\%), a post-war boom (4.35\%), and a return from 1977 to the pre-war rate (3.26\%). Mabe proposed to interpret this pattern as the succession of three epochs of science: ``innocent science'', ``big science'', and ``disillusionment science''. He suggested that the extraordinary resource inputs represented by the post Second World War funding temporarily accelerated growth before the system returned to its intrinsic equilibrium rate.

\citet{Gu_Blackmore} extended the analysis into the 21st century, using again Ulrich's data, and introducing some concepts of journal demographics (newly created journals, active, unpublished etc.). Their findings for the years 1986-2013 indicate a sustained high average growth rate (4.7\%), comparable to the mid-century ``big science'' boom. This expansion is characterized by a shift from print to online-only formats and the rise of Open Access. They suggested that the digital era has catalyzed a second major boom, fundamentally transforming the scholarly publishing system.

In describing and rationalizing journal demographics, \citet{Zsindely_Schubert} proposed applying demographic principles to journal populations. They established a direct analogy to human society: a journal population, like a human one, is composed of individuals—in this case, journals. Within this framework, launching a new journal equates to a birth and ceasing a journal to a death. Based on this demographic scheme, \citet{Jeannin} proposed specific measures for birth and death of journals. However, these proposals have remained largely isolated within the scientometric community.
 
There is a wealth of literature on editorial boards and their role, a significant portion of which is focused on the functioning and organization of journal peer review, also in a historical perspective \citep{Baldwin}. This literature is partly descriptive, seeking to understand the system's operations, and partly normative, aimed at suggesting improvements to it \citep{Horbach, Kochetkov}. More central from our point of view is the literature focused on the composition of editorial boards. Here, the literature is stretched between two poles: the descriptive and the evaluative. This tension began with \citet{zsindely1982}, one of the very first articles, which proposed using editorial board data for scientometric studies and examined the correlations between the number of editors of a journal and other bibliometric indicators. 

This literature originated with \citet{crane1967}, which explicitly characterized editorial board members as scientific ``gatekeepers''. Crane credited \citet{degrazia1963} with first applying the term ``gatekeepers''. Indeed, \citet{degrazia1963} used the term to refer to scholars within the ``reception system of science'' who are empowered to determine which pieces of knowledge are accepted as established and which are rejected. Moving beyond the view that Mertonian norms solely govern the publishing system, \citet{crane1967} emphasized the significant influence of the sociological composition of editorial boards on their decisions. By analyzing three journals, she tested the real-world application of these norms in peer review. Her findings revealed a shared academic profile among contributors and editors, including institutional affiliation, doctoral origin, and career stage. She concluded that this homogeneity persisted even with anonymous review, suggesting that social characteristics unavoidably shape editorial outcomes and that practice departs from Mertonian principles. 

Subsequent research  has largely focused on board composition. The editorial board have been analyzed from the point of view of their geographical composition, based mainly on affiliations of members  \citep{zsindely1982, hodgson1999, braun2005a, braun2005b, leydesdorff2009, baccini_re}. These analysis are used to identify the concentration of members in certain areas, notably the United States, Canada, and Western Europe; the lower participation from low income countries \citep{xu_et_al} and also to build indicators of internationalization of disciplines \citep{mazov2016}. Affiliations of members are also used to evaluate the institutional diversity of journals \citep{wu2020, baccini_re}. Studies on gender composition have documented the under-representation of women, especially in Editor-in-Chief roles and in the most prestigious journals \citep{hatfield1995,  addis2003, stegmaier2011, mauleon2013, metz2016, mazov2016, Feeney, baccini_re}. 

\citet{openeditors} developed the Open Editors project to address the critical lack of open, structured data on academic journal editors. By automatically scraping publicly available information from more than 7,000 journals across 26 publishers, the project created a unified dataset of approximately 600,000 researchers. This freely accessible dataset reveals significant heterogeneity in editorial board sizes, geographic diversity, and role labels. Unfortunately, it appears to be limited to a single snapshot, with data pertaining to 2022.

A stream of literature analyzes the network structures created by editors serving on multiple editorial boards. The concept of “interlocking editorships” was introduced by \citet{baccini2009} and was initially applied to examine relationships between journals in economics \citep{baccini2010}, statistics \citep{bacciniet2009}, and library and information science \citep{baccini2011}. This framework is used to map scientific communities (“invisible colleges”) and identify influential academic gatekeepers. A parallel line of inquiry was pioneered by \citet{sharma}. Subsequent research has been taken up within the scientometric community \citep[among the others, ][]{ni2010, Ni2013, andrikopoulos2015,teixeira2018, mendonca, goyanes2020, csomos2022, de_marcos,Goyanes_2025}. Recent developments focus on the relative contribution of interlocking editorships in defining journal similarity and clusters, compared to other dimensions such as shared authors, shared references, and textual similarities \citep{Baccini_Gingras, baccini_sim_net_fus, baccini_shape_2025}.

The extant literature on this subject adopt a static approach, in that analyses are conducted for a single point in time, both with regard to board composition and interlocking editorship, without consideration of temporal dynamics. This is probably largely due to the considerable difficulty of obtaining longitudinal data on editorial boards, which historically has required laborious manual collection. More recently developed web scraping techniques have not yet proven entirely effective for large-scale historical reconstruction  \citep{openeditors}. 
Consequently, little is known about the evolution of editorial boards over time. Furthermore, the analysis of editorial boards is completely disconnected from the demographics of journals. This article aims to address this gap by defining a framework for the dynamic study of editorial boards. It investigates the relationship between journal demographics, the dynamics of seats (positions) on their editorial boards, and the dynamics of the members who occupy these seats, thereby offering a longitudinal perspective.

\section{A stock-flow model for the dynamics of journal editorial boards}

The proposed stock-flow framework for analyzing the dynamics of journals and their editorial boards adopts the scholarly journal as its basic unit of observation. Each uniquely identified journal is assumed to be observed annually, resulting in a complete yearly list of editorial board members for all active journals. Furthermore, each member is uniquely and unambiguously identified in all the years in which it has been observed.

The demographic framework for journals is constructed by analyzing the differences between the populations of active journals from one year to the next.

The list of editorial board members of journals can be analyzed from two complementary perspectives: one focusing on the positions or seats within the boards, and the other on the scholars occupying those seats. Adopting the first perspective allows us to examine changes in board size over time, as well as the creation and elimination of seats at the journal level and other aggregated levels. The second perspective enables an analysis of compositional changes in board membership across time. 

The following subsections formalize the stock-flow framework that connects journal demographics to editorial board dynamics, examining both the available seats and the scholars who fill them.

\subsection{Journal stocks and flows}
Let $\mathbb{J}_t$ denote the set of active journals in year $t$,  with $J_t = |\mathbb{J}_t|$ being the total number of journals.

Let $\mathbb{N}_t \subseteq \mathbb{J}_t$ denote the set of newly created journals that began publication in year $t$, with $|\mathbb{N}_t| = N_t$. 

The journals that ceased publication between $t-1$ and $t$ form the set $\mathbb{X}_t \subseteq \mathbb{J}_{t-1}$, with cardinality $|\mathbb{X}_t| = X_t$.

The journals active in both year $t$ and $t-1$ form the set of persistent journals $\mathbb{P}_t = \mathbb{J}_t \cap \mathbb{J}_{t-1}$, with cardinality $|\mathbb{P}_t| = P_t$.

The total number of journals active in year $t$ can be represented as the algebraic sum of the previous year's stock ($J_{t-1}$), augmented by $N_t$, the new journals in year $t$, and diminished by $X_t$, the journal exits during the intervening period:
\begin{equation*}
J_t = J_{t-1} + N_t - X_t,
\label{number_journals}
\end{equation*}
and obviously $J_t = P_{t} + N_t$.

To contextualize entry and exit flows relative to the size of the journal population, we convert absolute numbers into rates. Since the population changes between $t-1$ and $t$, we use the average stock as the appropriate denominator:
\begin{equation*}
 \bar{J}_t = \frac{J_t + J_{t-1}}{2}.
\end{equation*}

The use of the average stock ensures rate symmetry between positive and negative changes. The standard growth rate of journals between $t-1$ and $t$ is usually defined as
\begin{equation*}
    \dot{J}^s_t = \frac{J_t - J_{t-1}}{J_{t-1}}
\end{equation*}
with asymmetric bounds: $\dot{J}^s_t \in [-1, \infty)$.

The symmetric \textit{journal net growth rate} is defined as:
\begin{equation}
\dot{J}_t = \frac{J_t - J_{t-1}}{\bar{J}_t}
\label{eq:journal_growth_rate}
\end{equation}
and ensures symmetric bounds: $\dot{J}_t \in [-2, 2]$. 
The two rates are related through the following transformation:
\begin{equation*}
\dot{J}_t = \frac{2\dot{J}^s_t}{1 + \dot{J}^s_t}
\quad \text{and} \quad 
\dot{J}^s_t = \frac{\dot{J}_t}{2 - \dot{J}_t}.
\end{equation*}

The \textit{journal creation rate} measuring the inflow of new journals, is defined as:
\begin{equation}
    \dot{N}_t = \frac{N_t}{\bar{J}_t};
    \label{eq:journal_entry_rate}
\end{equation}
the \textit{journal destruction rate}, measuring the exit of journals, is defined as:
\begin{equation}
    \dot{X}_t = -\frac{X_t}{\bar{J}_t};
    \label{eq:journal_exit_rate}
\end{equation}
the \textit{journal persistence rate} is:
\begin{equation}
    \dot{P}_t = \frac{P_t}{\bar{J}_t};
    \label{eq:journal_persistence_rate}
\end{equation}
and, finally, the \textit{journal turnover rate} is
\begin{equation}
    \dot{T}_t = \frac{N_t + X_t}{\bar{J}_t};
    \label{eq:journal_turnover_rate}
\end{equation}

The net journal growth rate can be decomposed as:
\begin{equation*}
    \dot{J}_t = \dot{N}_t + \dot{X}_t.
\end{equation*}

\subsection{Seat stocks and flows}

We define a ``seat'' as an editorial board position occupied by a scholar. The term ``seat'' is preferred over ``position'' to emphasize that we count only filled positions, excluding vacant ones. This analysis does not distinguish between roles (e.g., editor-in-chief, member, managing editor) and treats all seats as equivalent. All measurements of the stock of seats, of seat creation and seat destruction are derived from the cumulative aggregation of journal-level data. Aggregation may be performed at various levels, such as the entire set of journals, specific journal groups, or distinct subfields. 

The basic unit of analysis is the size of the editorial board of the journal $j$ at time $t$, defined as the number of seats on its board and denoted as $s_{j,t}$.

The variation of the board size of journal $j_{i,t}$ between time $t-1$ and $t$ is defined as 
\begin{equation*}
    \Delta s_{j,t}=s_{j,t}-s_{j,t-1}.
\end{equation*}

The rate of growth of the board of journal $j$ between $t$ and $t-1$ is denoted as $\dot{g}_{j,t}$ and computed in reference to the average size of journal board in the two considered years. The average size is defined as: 
\begin{equation*}
    z_{j,t}=\frac{s_{j,t}+ s_{j,t-1}}{2}
\end{equation*}
and the \textit{journal board growth rate} is:
\begin{equation}
    \dot{g}_{j,t}=\frac{s_{j,t}-s_{j,t-1}}{z_{j,t}}.
    \label{eq:journal_board_growth}
\end{equation}
As previously seen, the use of average size bounds $\dot{g}_{j,t}$ symmetrically between $-2$ and $+2$. For instance, a start-up journal has a growth rate of $+2$, and a discontinued journal has a growth rate of $-2$.

The total stock of editorial board seats at time $t$, denoted by $S_t$, 
is obtained by summing the board sizes of all active journals:
\begin{equation*}
S_t = \sum_{j \in \mathbb{J}_t} s_{j,t}.
\label{eq:seat_total}
\end{equation*}

In order to calculate aggregate seat stocks and flows, we consider four distinct sets of journals: two generating positive seat flows and two generating negative flows.

The first set contains persistent journals publishing in both $t-1$ and $t$ that expanded their board size between the two years; it is denoted as $\mathbb{P}^+_t \subseteq \mathbb{P}_t$ with cardinality $|\mathbb{P}^+_t| = P^+_t$.

These expanding journals generate a positive flow of seats calculated as:
\begin{equation}
C^{+}_t = \sum_{j \in \mathbb{J}^{+}} (s_{j,t} - s_{j,t-1}).
\label{eq:seat_creation_expanding}
\end{equation}

The second set consists of newly established journals \( \mathbb{N}_t \), with cardinality $\mathbb|{N}_t|=N_t$,  which generate a positive flow of seats:
\begin{equation}
C^{\mathbb{N}}_t = \sum_{j \in \mathbb{N}_t} s_{j,t}.
\label{eq:seat_creation_new}
\end{equation}

The third set, denoted as $\mathbb{P}^{-}_t \subseteq \mathbb{P}_t$ with cardinality $\mathbb|{P}^-_t|=P^-_t$, contains the contracting journals, that is, the persistent journals publishing both at $t-1$ and $t$ that contracted their board size between the two years. These contracting journals create a negative flow of seats calculated as: 
\begin{equation}
D^{-}_t = \left| \sum_{j \in \mathbb{J}^{-}} (s_{j,t} - s_{j,t-1}) \right|.
\label{eq:seat_destruction_contracting}
\end{equation}

The fourth set \( \mathbb{X}_t \) consists of discontinued journals between $t-1$ and $t$ which generate a negative flow of seats:
\begin{equation}
D^{\mathbb{X}}_t = \sum_{j \in \mathbb{X}_t} s_{j,t}.
\label{eq:seat_destruction_exit}
\end{equation}

A fifth set, denoted $\mathbb{P}^S_t\subseteq \mathbb{P}_t$ with cardinality $|\mathbb{P}^S_t|=P^S_t$, comprises persistent journals that maintained a constant board size between $t$ and $t-1$. 

Hence, the total number of seats of the journals active in year $t$ can be represented as the algebraic sum of the previous year's stock of seats ($S_{t-1}$), augmented by the new seats generated by new journals and by expanding journals, and diminished by the seats destroyed due to journals contracting their boards and discontinuation during the intervening period:
\begin{equation*}
S_t = S_{t-1} + C^{+}_t + C^{\mathbb{N}}_t - D^{-}_t - D^{\mathbb{X}}_t.
\end{equation*}

The \textit{Gross Seat Creation} at time $t$, denoted as $C_t$ is defined as the sum of seat gains across all journals that either expanded their board size or were newly established between $t-1$ and $t$:
\begin{equation}
C_t = C^{+}_t + C^{\mathbb{N}}_t =\sum_{j \in \mathbb{J}^{+}} (s_{j,t} - s_{j,t-1})+ \sum_{j \in \mathbb{N}_t} s_{j,t}.
\label{eq:gross_seat_creation}
\end{equation}

The \textit{Gross Seat Destruction} at time $t$, denoted as $D_t$, is defined as the sum of seat losses across all journals that either reduced their board size or were discontinued between $t-1$ and $t$:
\begin{equation}
D_{t} = D^{-}_t + D^{\mathbb{X}}_t = \left|\sum_{j \in \mathbb{J}^{-}} (s_{j,t} - s_{j,t-1})\right| + \sum_{j \in \mathbb{X}_t} s_{j,t}.
\label{eq:gross_seat_destruction}
\end{equation}
The \textit{net aggregate seat change} at time $t$ for the set of journals $\mathbb{J}_t$, denoted $\Delta{S}_{t}$, is the difference between the total number of seats of all journals at time $t$ and the total number of seats at time $t-1$:
\begin{equation}
    \Delta S_{t}=S_{t}-S_{t-1}.
    \label{eq:seat_change}
\end{equation}
Journals with an unchanged number of seats on their boards contribute neither to seat creation nor to seat destruction. Hence, 
\begin{equation*}
    \Delta S_{t}=C_{t}-D_{t}.
\end{equation*}
The seat creation and destruction decompose the net seat change by providing information about board dynamics. 

The \textit{gross seat turnover} indicates the total number of seat reallocated at time $t$, and is defined as
\begin{equation*}
    T^S_{t}=C_{t}+D_{t}.
    \label{seat_reallocation}
\end{equation*}

These aggregate measures can be converted into rates by dividing them by a measure of the size of the journal boards. Also in this case this aggregate measure of size, denoted by $Z_{t}$, is computed as the average of the total number of seats of all journals between $t$ and $t-1$:
\begin{equation*}\
    Z_{t}= \frac{S_{t}+ S_{t-1}}{2}.
\end{equation*}

The \textit{rate of seat creation} is defined as
\begin{equation}
\dot{C}_{t} = \frac{C_{t}}{Z_{t}}.
\label{eq:seat_creation_rate}
\end{equation}

It can be decomposed into the rate generated from expanding journals and from new journals. They are defined respectively as:
\begin{equation*}
\dot{C}^+_{t}=\frac{C^{+}_{t}}{Z_{t}}   
\label{eq:seat_creation_rate_expanding}
\end{equation*}
and
\begin{equation*}
\dot{C}^\mathbb{N}_{t}= \frac{C^{\mathbb{N}}_{t}}{Z_{t}};
\label{eq:seat_creation_rate_new}
\end{equation*}
by construction:
\begin{equation*}
\dot{C}_{t} = \dot{C}^+_{t}+\dot{C}^\mathbb{N}_{t}.
\end{equation*}

Analogously, the \textit{rate of seat destruction} is defined as
\begin{equation}
\dot{D}_{t} = -\frac{D_{t}}{Z_{t}}. 
\label{eq:seat_destruction_rate}
\end{equation}

The rate of seat destruction generated by contracting journals is:
\begin{equation}
\dot{D}^-_t = -\frac{D^{-}_{t}}{Z_{t}}, 
\label{eq:seat_destruction_rate_contracting}
\end{equation}
and the rate generated by discontinued journals is 
\begin{equation}
\dot{D}^X_t = - \frac{D^\mathbb{X}_t}{Z_{t}},
\label{eq:seat_destruction_rate_exit}
\end{equation}
and obviously:
\begin{equation*}
\dot{D}_{t} = \dot{D}^-_t + \dot{D}^X_t .
\end{equation*}

The \textit{aggregate rate of seat growth} for all the journals is 
\begin{equation}
    \dot{S}_{t} = \frac{S_t - S_{t-1}}{Z_t} = \frac{C_t-D_t}{Z_t} = \dot{C}_t + \dot{D}_t,
    \label{eq:seat_aggregate_growth_rate}
\end{equation}
and the complete decomposition of the rate is therefore 
\begin{equation*}
    \dot{S}_{t} = \frac{C^{+}_{t}}{Z_{t}} + \frac{C^{\mathbb{N}}_{t}}{Z_{t}} + \frac{D^{-}_{t}}{Z_{t}} + \frac{D^{\mathbb{X}}_{t}}{Z_{t}}.
\end{equation*}

Finally, the \textit{total seat turnover rate} is defined as:
\begin{equation}
    \dot{T}^S_t=\dot{C}_t-\dot{D}_t=\frac{C_t}{Z_t}-\frac{D_t}{Z_t}.
    \label{eq:seat_reallocation_rate}
\end{equation}

All these growth rates are symmetric and easily comparable since they are based on the same denominator $Z_t$ and are bounded in the interval $[-2, +2]$.

\subsection{Member stocks and flows}

The analysis of board seats does not account for the individual scholars who occupy them. It is therefore useful to examine journal governance by focusing on scholars rather than on the number of available positions.

Stock-flow analysis can be conducted at two different levels of aggregation. The more aggregate approach considers the entire set of editorial board members as a whole and studies their overall stock and flows. In contrast, a journal-level analysis examines stock and flows for each journal individually.

The relationship between these two levels is not straightforward. Consider the case of a ``new member''. At the aggregate level, a new member in a given year is a scholar who joins the board of one or more journals and was not present on any editorial board in the previous year. At the individual journal level, however, a new member is a scholar joining that specific journal's board in that year, regardless of whether they already served on other journals' boards in the previous year. Naturally, some journal-level new members are also new members at the aggregate level. 
A parallel distinction applies to scholars leaving editorial boards. At the journal level, an exit refers to a scholar departing from that specific journal's board, while at the aggregate level, it denotes a scholar exiting the entire editorial system (i.e., holding no editorial positions in the given year despite having held at least one in the previous year).

To operationalize these distinctions, we develop a stock-flow framework that separates the aggregate from the journal-level dynamics.

Let \( \mathbb{M} \) denote the set of all scholars who served on the editorial board of at least one journal during the entire period under examination. 
The total number of scholars who served as editors at least once during the entire period is given by $M = |\mathbb{M}|$. 

The set of editorial board members active in year $t$ is denoted by $\mathbb{M}_t \subseteq \mathbb{M}$, and the number of active members in the year $t$ is $M_t = |\mathbb{M}_t|$.

The set of scholars retained from year $t-1$ to $t$ is $\mathbb{M}^R_t = \mathbb{M}_t \cap \mathbb{M}_{t-1}$, with cardinality $M^R_t = |\mathbb{M}^R_t|$. 
The set of new scholars in year $t$ is $\mathbb{M}^N_t = \mathbb{M}_t \setminus \mathbb{M}_{t-1}$, with cardinality $M^N_t = |\mathbb{M}^N_t|$. 
Finally, the set of scholars who served on at least one editorial board in year $t-1$ but were not present on any board in year $t$ is $\mathbb{M}^X_t = \mathbb{M}_{t-1} \setminus \mathbb{M}_t$, with cardinality $M^X_t = |\mathbb{M}^X_t|$.

The stock of active members at year $t$ is therefore computed as:
\begin{equation*}
    M_t=M_{t-1}+M_t^N-M^X_t,
    \label{stock_members_aggregate}
\end{equation*}
and obviously: 
\begin{equation*}
M_t=M^R_{t}+M_t^N.
\end{equation*}
The \textit{net aggregate change of members} at time $t$ is the difference between the total number of members at time $t$ and the total number of members at $t-1$:
\begin{equation*}
    \Delta{M}_t=M_t-M_{t-1}=M_t^N-M^X_t.
    \label{stock_members_aggregate_1}
\end{equation*}

As previously done for journals and seats, flow rates are computed by considering the average size of the board members population between $t-1$ and $t$ as denominator: 
\begin{equation*}
\bar{M}_t = \frac{M_t + M_{t-1}}{2}.
\end{equation*}

The \textit{growth rate of members} between year $t-1$ and $t$ is computed as:
\begin{equation}
    \dot{M}_t = \frac{M_t - M_{t-1}}{\bar{M}_t}
    \label{eq:member_growth_rate}
\end{equation}
and is bounded in the interval $[-2, 2]$.

The \textit{entry rate of new members} is defined as:
\begin{equation}
    \dot{M}^N_t = \frac{M^N_t}{\bar{M}_t}
    \label{eq:member_entry_rate}
\end{equation}
and is bounded in $[0, 2]$.

The \textit{exit rate of members} who become inactive is defined as:
\begin{equation}
    \dot{M}^X_t = -\frac{M^X_t}{\bar{M}_t}
    \label{eq:member_exit_rate}
\end{equation}
and is bounded in $[-2, 0]$.

The \textit{member turnover rate} is defined as:
\begin{equation}
    \dot{T}^M_t = \frac{M^N_t+M^X_t}{\bar{M}_t}
    \label{eq:member_turnover_rate}
\end{equation}
and is bounded in $[0, 2]$.

The \textit{member retention rate}, defined as the proportion of scholars serving on editorial boards in both year $t$ and $t-1$, is:
\begin{equation}
    \dot{M}^R_t = \frac{M^R_t}{\bar{M}_t}
    \label{eq:member_retention_rate}
\end{equation}
and is bounded in $[0, 1]$.

By combining information about seats with information about members, it is possible to compute for year $t$ the share of gross seat creation covered by new members. The new member coverage ratio is defined as:
\begin{equation}
\rho_t = 
\begin{cases}
\frac{M^N_t}{C_t} & \text{if } C_t > 0 \\
0 & \text{if } C_t = 0
\end{cases}.
\label{eq:new_member_coverage}
\end{equation}

The journal-level analysis builds upon the overall set of unique members $\mathbb{M}$. 
The editorial board composition of journal $j$ at time $t$ is defined as the set of scholars serving on its board: $\mathbb{M}_{j,t} \subseteq \mathbb{M}$, with cardinality $m_{j,t} = |\mathbb{M}_{j,t}|$. 

The set of scholars retained in journal $j$ from year $t-1$ to $t$ is $\mathbb{M}^R_{j,t} = \mathbb{M}_{j,t} \cap \mathbb{M}_{j,t-1}$, with $m^R_{j,t} = |\mathbb{M}^R_{j,t}|$. 

The subset of these who are new in journal $j$ and completely new to the editorial system, that is not present on any editorial board at $t-1$, are denoted as:  $\mathbb{M}^N_{j,t} \subseteq \mathbb{M}^n_{j,t}$, with $m^N_{j,t} = |\mathbb{M}^N_{j,t}|$. 

The scholars who exited a journal $j$ between $t-1$ and $t$ are 
$\mathbb{M}^x_{j,t} = \mathbb{M}_{j,t-1} \setminus \mathbb{M}_{j,t}$, 
with $m^x_{j,t} = |\mathbb{M}^x_{j,t}|$.

Finally, scholars who left journal $j$ at time $t$ while simultaneously exiting 
the entire editorial system are denoted $\mathbb{M}^X_{j,t} \subseteq \mathbb{M}^x_{j,t}$, 
with $m^X_{j,t} = |\mathbb{M}^X_{j,t}|$.

The stock of members in journal $j$ at year $t$ is therefore computed as:
\begin{equation*}
    m_{j,t} = m_{j,t-1} + m_{j,t}^n - m^x_{j,t},
    \label{eq:member_journal}
\end{equation*}
and obviously:
\begin{equation*}
    m_{j,t} = m^R_{j,t} + m_{j,t}^n.
\end{equation*}

As previously done, rates for journal $j$ are calculated by using the average number of its members  between $t-1$ and $t$ as denominator: 
\begin{equation*}
\bar{m}_{j,t} = \frac{m_{j,t} + m_{j,t-1}}{2}.
\end{equation*}

The growth rate of members for journal $j$ between year $t-1$ and $t$ is computed as:
\begin{equation}
    \dot{m}_{j,t} = \frac{m_{j,t} - m_{j,t-1}}{\bar{m}_{j,t}}
    \label{eq:member_growth_journal}
\end{equation}
and is bounded in the interval $[-2, 2]$. The two extreme cases are verified respectively when a journal is discontinued and loses all its previous members, and when a new journal starts its publication by defining a board where all members are obviously new. 

The entry rate of members who are new for a journal $j$ is defined as:
\begin{equation}
    \dot{m}^n_{j,t} = \frac{m^n_{j,t}}{\bar{m}_{j,t}}
    \label{eq:new_member_rate_journal}
\end{equation}
and is bounded in $[0, 2]$; analogously the entry rate of members who are completely anew for the editorial system is defined:
\begin{equation}
    \dot{m}^N_{j,t} = \frac{m^N_{j,t}}{\bar{m}_{j,t}}.
    \label{eq:New_member_rate_journal}
\end{equation} 

The member exit rate for journal $j$ is defined as:
\begin{equation}
    \dot{m}^x_{j,t} = -\frac{m^x_{j,t}}{\bar{m}_{j,t}}
    \label{eq:exit_member_rate_journal}
\end{equation}
and is bounded in $[-2, 0]$.

The member turnover for journal $j$ is defined as the proportion of scholars who enter in or exit from its editorial board between year $t$ and $t-1$:
\begin{equation}
    \dot{T}^m_{j,t} = \frac{m^n_{j,t}+m^x_{j,t}}{\bar{m}_{j,t}}
    \label{eq:turnover_member_rate_journal}
\end{equation}

The member retention rate for journal $j$ is defined as the proportion of scholars serving on its editorial board in both year $t$ and $t-1$:
\begin{equation}
    \dot{m}^R_{j,t} = \frac{m^R_{j,t}}{\bar{m}_{j,t}}
    \label{eq:member_retention_rate_journal}
\end{equation}
and is bounded in $[0, 1]$.

These rates are linked by two demographic identities:
\begin{equation*} 
\dot{m}_{j,t} = \dot{m}^n_{j,t} + \dot{m}^x_{j,t}, 
\end{equation*}
\begin{equation*}
\dot{m}^R_{j,t} + \dot{m}^n_{j,t} = \frac{m_{j,t}}{\bar{m}_{j,t}}
\end{equation*}
reflecting how net growth decomposes into incoming and outgoing flows, with retention representing the stable core of the editorial board.

By combining information about seats and members, it is possible to compute for journal $j$ and year $t$ the share of seat creation covered by members who are new for the editorial system. The new member coverage ratio for journal $j$ is defined as:
\begin{equation}
\rho_{j,t} = 
\begin{cases}
\frac{m^N_{j,t}}{\Delta s_{j,t}} & \text{if } \Delta s_{j,t} > 0 \\
0 & \text{if } \Delta s_{j,t} = 0
\label{eq:coverage_journal}
\end{cases}.
\end{equation}

\subsection{Adjustment for data observed in different time-intervals} 

In the previous two subsections, we defined the general framework for studying stock-flow analysis within defined time periods, where all flows are measured between $t$ and $t-1$. However, when observations and flows span different intervals, data adjustments become necessary.  In our case, for example, data are observed every ten years from 1866 to 2006, but the final two observations -- 2012 and 2019 -- refer to a 6-year and 7-year period respectively. 
In these shorter intervals, journals have had less time to adjust their trajectories. Consequently, observed growth rates may reflect either genuine structural changes or simply the compressed timeframe for adjustment.

The standard solution is to annualize the rates by dividing the total change by the number of years in the observation period. This solution has a strong shortcoming that can be illustrated with an example. As we have seen, a new founded journal has a rate of growth of $2$, and a discontinued journal of $-2$. Dividing each rate by the number of years in its observation period produces annualized rates with different ranges across periods: the annual growth rate for new (discontinued) journals would be $0.2$ ($-0.2$) in a ten-year period, $0.333$ ($-0.333$) in a six-year period, and $0.286$ ($-0.286$) in a seven-year period.

To overcome this problem, we adjust annualized rates by normalizing them according to their maximum possible value, given by $2/P_t$, where $P_t$ is the number of years in the observation period. In this manner, the annualized rates are defined in the interval $[-1, +1]$.

Consider, for example and for all, the aggregate seat growth of boards $\dot{S}_{t}$; it can be annualized as:
\begin{equation*}
\dot{S}^y_{t} = \frac{\dot{S}_t}{P_t} = \frac{S_t-S_{t-1}}{Z_{t} \times P_t}.
\end{equation*}

The normalized annual rate is then computed by dividing the rates by the maximum value of the rate in the considered period:  
\begin{equation*}
\dot{S}^{ny}_{t} = \frac{S_t-S_{t-1}}{Z_{t} \times P_t} \times \frac{P_t}{2}
\end{equation*}
which, recalling that $Z_t = \frac{S_t + S_{t-1}}{2}$, simplifies to:
\begin{equation*}
\dot{S}^{ny}_{t} = \frac{S_t-S_{t-1}}{2Z_{t}} = \frac{S_t-S_{t-1}}{S_t + S_{t-1}}.
\end{equation*}

This statistic captures net growth or decline as a proportion of the total number of seats that existed across the two periods. A value of $+0.3$, for example, indicates that 30\% of all positions present in either $t-1$ or $t$ represent net new additions. To illustrate, if the number of seats increases from $S_{t-1}=400$ to $S_t=600$, the normalized rate is
\[
\frac{600 - 400}{600 + 400} = \frac{200}{1000} = 0.2,
\]
meaning that 20\% of the combined 1{,}000 seats across the two periods reflects net expansion.

This formulation expresses the intensity of change relative to the overall scale of the editorial board system, ensuring a better comparability across observation periods of different durations by eliminating mechanical period-length effects.

This normalization may be satisfactory when applied to the demographics of journals and to seat dynamics. The dynamics of the members may be more problematic. Due to their academic and natural longevity, scholars likely tend to enter and exit journal editorial boards in the latter part of the chosen observation period, regardless of its length. Hence, the assumption that annual averages are easily comparable when calculated over different time frames should be adopted with great caution. 

\section{Data}

The stock-flow framework developed in the previous section is applied to data from economics journals. The journals considered are those covered by the ``Gatekeepers of Economics Longitudinal Database'' (GOELD). As anticipated, the data were collected for one year per decade from 1866 to 2006, plus the years 2012 and 2019. This irregular periodization results from the project's extended timeline (2006–2025), due to the vagaries of funding availability for supporting the manual harvesting of data.

GOELD incorporates data from journals indexed in EconLit, based on a list compiled from the American Economic Association's website in April 2019. \footnote{\url{https://web.archive.org/web/20190716024210/https://www.aeaweb.org/econlit/journal_list.php}} According to \citet{Gusenbauer}, \textit{EconLit} provides the best bibliographic coverage of the economics literature and a limited coverage of the business literature. The choice of this list of journals aims to capture the boundaries of the field as self-defined by the primary scholarly society in economics. The result is a corpus that is more inclusive than those provided by major bibliographic databases such as Web of Science or Scopus. It includes publications from intersecting domains that the economic community actively engages with, but is primarily based on English-language sources, potentially underrepresenting journals from the Global South and critical interdisciplinary fields. This focus provides a coherent analytical framework centered on the community's own conception of its publishing domain.

In terms of size, GOELD comprises 1,724 journals, featuring 2,202 name variants resulting from title changes over time. The dataset includes 71,173 observed editorial boards, amounting to 206,217 seats occupied by 100,998 unique members.
Most members are assigned a gender, and, where available, their affiliations are geotagged, as described in more detail by \citet{baccini_re}.
Data were harvested manually from journal websites for the most recent observed year, and from electronic or hard-copy versions of journals for all preceding years. The disambiguation of editorial board member names was carried out in three phases. Due to the delayed availability of data for different years, each year was disambiguated manually as the data became available. Upon completion of the data collection, a longitudinal disambiguation was conducted in two phases: a search for string similarity, followed by manual disambiguation. 

The complete GOELD database will be released when the overarching research project—of which this article is a part—is completed. However, the anonimized data necessary to reproduce the results presented here are available from \url{https://10.5281/zenodo.18186430}.

\section{Data analysis}

In this study, we treat economics journals as an undifferentiated set. This choice is motivated by the extended temporal coverage of our analysis, which spans several decades. Journal rankings and impact measures are not consistently available for the entire period, and cannot be reliably reconstructed from existing data without sacrificing coverage of a substantial majority of journals \citep{Truc_2021}. Similarly, no systematic classification of journals exists for the time span under investigation \citep{baccini_shape_2025}. However, the methodology presented here is flexible and can be applied directly to the analysis of any subset of journals, regardless of their definition, whether based on discipline, prestige, type of production, or structural characteristics. 

\subsection{The demographics of economics journals (1866-2019)}

\begin{figure}
  \centering

  \begin{subfigure}[t]{0.90\textwidth}
    \centering
    \includegraphics[width=\linewidth]{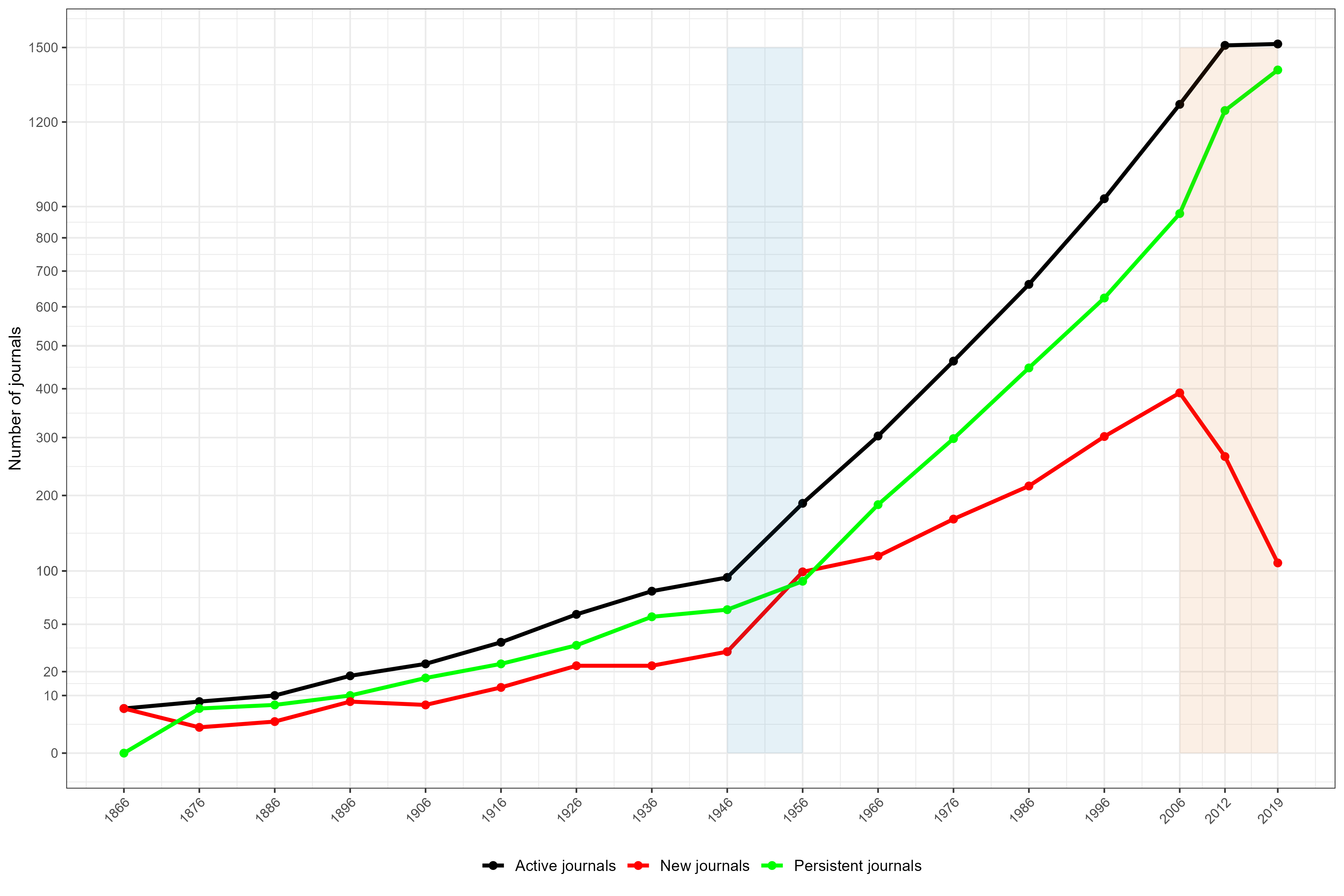}
    \caption{Total number of active, persistent and newly established economics journals (1866-2019).}
    \label{fig:panelA}
  \end{subfigure}
  \hfill
  \begin{subfigure}[t]{0.90\textwidth}
    \centering
    \includegraphics[width=\linewidth]{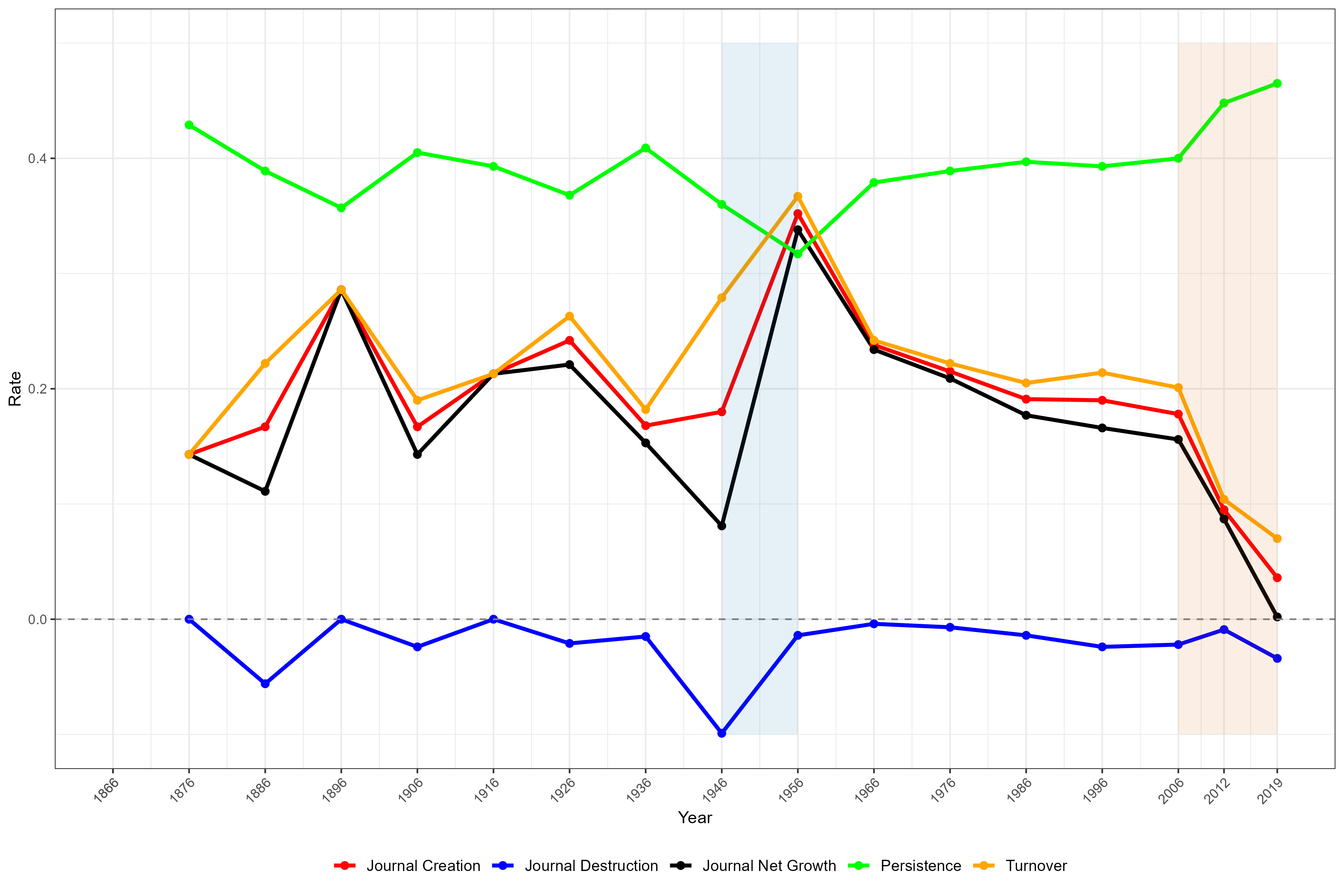}
    \caption{Normalized Rates of journal creation ($\dot{N}_t$, defined in eq. \ref{eq:journal_entry_rate}), destruction ($\dot{X}_t$, eq. \ref{eq:journal_exit_rate}), net growth ($\dot{J}_t$, eq. \ref{eq:journal_growth_rate}), persistence ($\dot{P}_t$, eq. \ref{eq:journal_persistence_rate}) and turn-over ($\dot{T}_t$, eq. \ref{eq:journal_turnover_rate}).}
    \label{fig:panelB}
  \end{subfigure}

  \caption{The demographics of economics journals (1866-2019).}
  \label{fig:journal_demographics}
\end{figure}

The demographic data about economics journals are illustrated in Figure \ref{fig:journal_demographics} and reported in Table \ref{tab:journal_demography} and Table \ref{tab:journal_rates}. 

The series can be partitioned into multiple phases. The first period ended with the shock of the Second World War (WWII). It marked the establishment of economics as an independent scientific field with the definition of dedicated research outlets and their governance structure. This phase is characterized by gradual consolidation, with the journal population growing from 6 to 93. The rate of journal creation was positive but volatile, while exit rates were generally low with the notable exception of the WWII decade. Consequently, turnover also exhibited volatility. Overall, in this first phase, the publishing environment of economics appears of small scale, slowly growing, and persistent: once founded, journals had a high probability of survival. 

\begin{table}
\caption{Demographics of economics journals (1866-2019). For each year: $J_t$ is the total number of active journals; $P_t$ counts persistent journals active in both $t$ and $t-1$ (with $P_t = P^+_t + P^S_t + P^-_t$); $N_t$ represents newly established journals; $X_t$ counts discontinued journals; $P^+_t$ tracks persistent journals expanding their board size; $P^S_t$ counts persistent journals with stable board size; $P^-_t$ counts persistent journals reducing their board size.} 
\label{tab:journal_demography}
\centering
\begin{tabular}[t]{rrrrrrrr}
\toprule
Year & $J_t$ & $P_t$ & $N_t$ & $X_t$ & $P^+_t$ & $P^s_t$ & $P^-_t$\\
\midrule
1866 & 6 & 0 & 6 & 0 & 0 & 0 & 0\\
1876 & 8 & 6 & 2 & 0 & 3 & 2 & 1\\
1886 & 10 & 7 & 3 & 1 & 3 & 3 & 1\\
1896 & 18 & 10 & 8 & 0 & 5 & 3 & 2\\
1906 & 24 & 17 & 7 & 1 & 9 & 5 & 3\\
\addlinespace
1916 & 37 & 24 & 13 & 0 & 8 & 10 & 6\\
1926 & 58 & 35 & 23 & 2 & 14 & 8 & 13\\
1936 & 79 & 56 & 23 & 2 & 20 & 18 & 18\\
1946 & 93 & 62 & 31 & 17 & 29 & 15 & 18\\
1956 & 188 & 89 & 99 & 4 & 48 & 16 & 25\\
\addlinespace
1966 & 303 & 186 & 117 & 2 & 100 & 40 & 46\\
1976 & 463 & 298 & 165 & 5 & 190 & 31 & 77\\
1986 & 662 & 447 & 215 & 16 & 292 & 36 & 119\\
1996 & 926 & 624 & 302 & 38 & 372 & 37 & 215\\
2006 & 1,268 & 877 & 391 & 49 & 509 & 62 & 306\\
\addlinespace
2012 & 1,509 & 1,244 & 265 & 24 & 744 & 118 & 382\\
2019 & 1,515 & 1,406 & 109 & 103 & 871 & 119 & 416\\
\addlinespace
0619 & 1,515 & 1,167 & 348 & 101 & 794 & 41 & 332\\
\bottomrule
\end{tabular}
\end{table}

The year 1956 emerges as a turning point: in the decade following WWII, the number of journals more than doubled, recording the highest net growth and turnover rates of the entire period. 

\begin{table}
\caption{Rates of journal creation ($\dot{N}_t$, defined in eq. \ref{eq:journal_entry_rate}), destruction ($\dot{X}_t$, eq. \ref{eq:journal_exit_rate}), net growth ($\dot{J}_t$, eq. \ref{eq:journal_growth_rate}), persistence ($\dot{P}_t$, eq. \ref{eq:journal_persistence_rate}) and turn-over ($\dot{T}_t$, eq. \ref{eq:journal_turnover_rate}) for the period 1866-2019. The row ``0619'' reports values computed over the 13-year interval 2006-2019. All rates are normalized.}
\label{tab:journal_rates}
\centering
\begin{tabular}[t]{cccccc}
\toprule
Year & $\dot{N}_t$ & $\dot{X}_t$ & $\dot{J}_t$ & $\dot{P}_t$ & $\dot{T}_t$\\
\midrule
1866 & 1.000 & 0.000 & 1.000 & 0.000 & 1.000\\
1876 & 0.143 & 0.000 & 0.143 & 0.429 & 0.143\\
1886 & 0.167 & -0.056 & 0.111 & 0.389 & 0.222\\
1896 & 0.286 & 0.000 & 0.286 & 0.357 & 0.286\\
1906 & 0.167 & -0.024 & 0.143 & 0.405 & 0.190\\
\addlinespace
1916 & 0.213 & 0.000 & 0.213 & 0.393 & 0.213\\
1926 & 0.242 & -0.021 & 0.221 & 0.368 & 0.263\\
1936 & 0.168 & -0.015 & 0.153 & 0.409 & 0.182\\
1946 & 0.180 & -0.099 & 0.081 & 0.360 & 0.279\\
1956 & 0.352 & -0.014 & 0.338 & 0.317 & 0.367\\
\addlinespace
1966 & 0.238 & -0.004 & 0.234 & 0.379 & 0.242\\
1976 & 0.215 & -0.007 & 0.209 & 0.389 & 0.222\\
1986 & 0.191 & -0.014 & 0.177 & 0.397 & 0.205\\
1996 & 0.190 & -0.024 & 0.166 & 0.393 & 0.214\\
2006 & 0.178 & -0.022 & 0.156 & 0.400 & 0.201\\
\addlinespace
2012 & 0.095 & -0.009 & 0.087 & 0.448 & 0.104\\
2019 & 0.036 & -0.034 & 0.002 & 0.465 & 0.070\\
\addlinespace
0619 & 0.125 & -0.036 & 0.089 & 0.419 & 0.161\\

\bottomrule
\end{tabular}
\end{table}

The five decades after this turning point saw an unprecedented demographic expansion. The number of journals exploded, reaching 1,268 by 2006 with a peak of 391 newly founded journals in the 1996-2006 decade. The expansionary phase was driven by a powerful combination of a persistently high rate of new journal creation, even if it had declined since its 1956 peak, minimal journal closures, and a gradually increasing persistence rate.

The period from 2006 to 2019, however, marks a distinct new phase. When considered over the entire 13-year span, the net growth rate of journals fell to its historical minimum, despite the longer timeframe compared to previous intervals. Both the exit rate and the persistence rate reached their highest levels of the expansive period that began in 1956. An analysis of the two subperiods reveals an even more pronounced shift. By 2019, the net growth rate had plummeted to near zero, while the persistence rate hits its historical peaks.

This convergence of journal creation and destruction rates in 2019 suggests that the economics discipline reached a state of saturation in publishing and a progressive closure to new journals. Furthermore, the unprecedented persistence rate indicates a field where established journals have become highly effective at insulating themselves from external pressures.

\subsection{Seats at the editorial tables of economics journals}

\begin{figure}[htbp]
    \centering
    \includegraphics[width=0.9\textwidth, height=0.6\textheight, keepaspectratio]{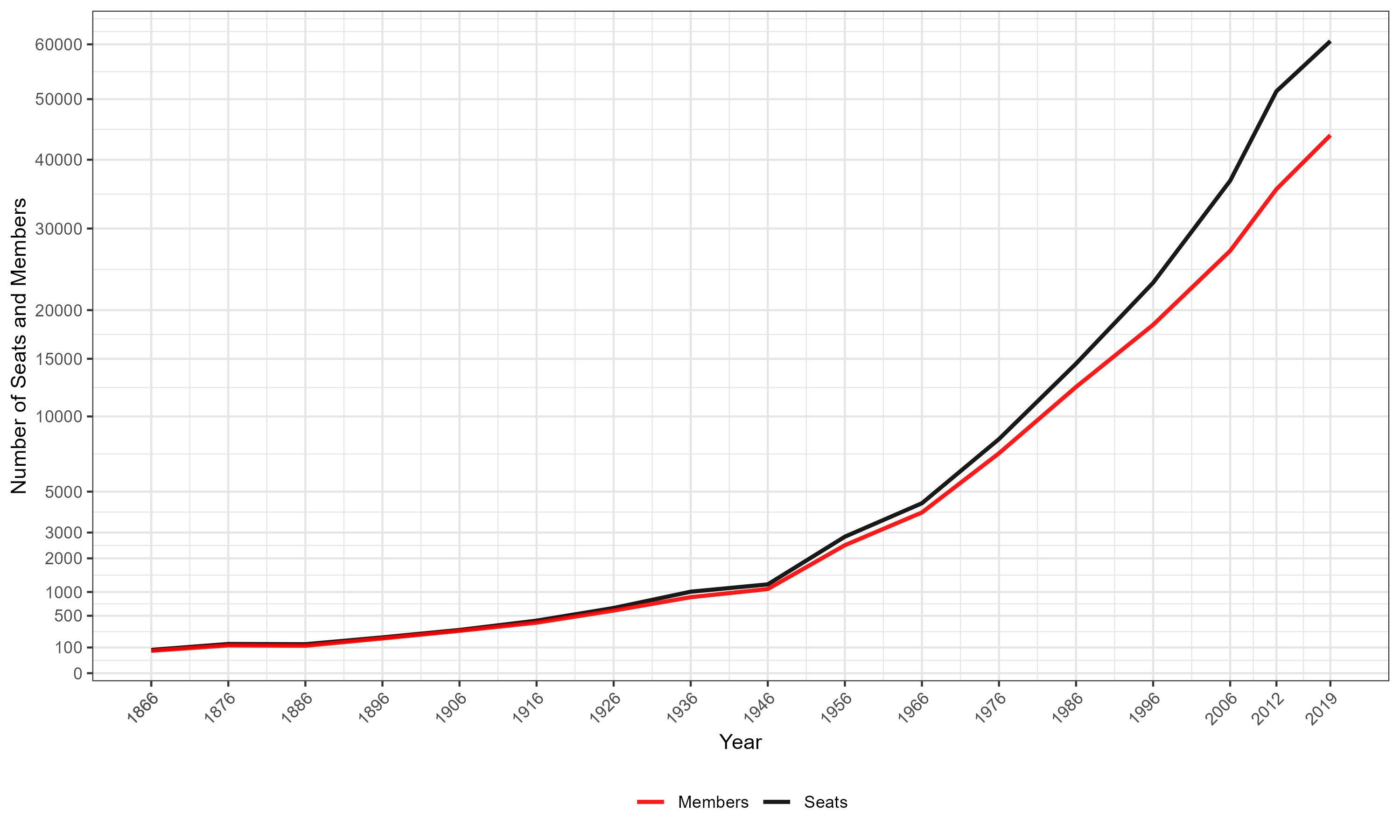}
    \caption{Number of seats and members in the editorial boards of Economics Journal (1866-2019)}
    \label{fig:seats_member_series}
\end{figure}

Stock data on editorial seats and members are reported in Table \ref{tab:seats_members_demography} and drawn in Figure \ref{fig:seats_member_series}. Although both the number of seats and the number of individual members grew substantially throughout the period, the expansion of editorial seats significantly outpaced the growth of scholars serving on boards. This divergence is the result of two related phenomena. First, the growing demand for editorial services has been largely met by scholars who were already active as editorial board members. Second, the increasing willingness of scholars to serve on multiple boards simultaneously reflects the rising importance of editorial roles, both as currency in academic careers and as a key mechanism of scholarly gatekeeping.

\begin{table}

\caption{Editorial governance of economics journals (1866-2019). Total seats ($S_t$), median board size ($\tilde{s}_t$) and average seats per journal ($S_t/J_t$). Total unique members ($M_t$), average members per journal ($M_t/J_t$) and average seats per member ($S_t/M_t$).}
\centering
\label{tab:seats_members_demography}
\begin{tabular}[t]{crrrrrr}
\toprule
Year & $S_t$ & $\tilde{s}_t$ & ${S_t}/{J_t}$ & $M_t$ & ${M_t}/{J_t}$ & ${S_t}/{M_t}$\\
\midrule
1866 & 82 & 5.5 & 13.67 & 76 & 12.67 & 1.08\\
1876 & 129 & 4.5 & 16.12 & 117 & 14.62 & 1.10\\
1886 & 127 & 4.0 & 12.70 & 114 & 11.40 & 1.11\\
1896 & 194 & 5.5 & 10.78 & 183 & 10.17 & 1.06\\
1906 & 283 & 8.0 & 11.79 & 272 & 11.33 & 1.04\\
\addlinespace
1916 & 418 & 7.0 & 11.30 & 387 & 10.46 & 1.08\\
1926 & 644 & 7.0 & 11.10 & 595 & 10.26 & 1.08\\
1936 & 1,007 & 5.0 & 12.75 & 876 & 11.09 & 1.15\\
1946 & 1,196 & 7.0 & 12.86 & 1,074 & 11.55 & 1.11\\
1956 & 2,825 & 8.0 & 15.03 & 2,484 & 13.21 & 1.14\\
\addlinespace
1966 & 4,381 & 10.0 & 14.46 & 3,918 & 12.93 & 1.12\\
1976 & 8,315 & 14.0 & 17.96 & 7,338 & 15.85 & 1.13\\
1986 & 14,541 & 19.0 & 21.97 & 12,433 & 18.78 & 1.17\\
1996 & 23,136 & 20.0 & 24.98 & 18,416 & 19.89 & 1.26\\
2006 & 36,805 & 24.0 & 29.03 & 27,075 & 21.35 & 1.36\\
\addlinespace
2012 & 51,383 & 28.0 & 34.05 & 35,550 & 23.56 & 1.45\\
2019 & 60,614 & 31.0 & 40.01 & 43,915 & 28.99 & 1.38\\
\bottomrule
\end{tabular}
\end{table}

\begin{table}

\caption{Seat creation ($C_t$, defined in eq. \ref{eq:gross_seat_creation}) and destruction ($D_t$, eq. \ref{eq:gross_seat_destruction}) in the period 1866-2019. Created seats comprise expansions of the boards of permanent journals ($C^+_t$, eq. \ref{eq:seat_creation_expanding}) and new journals ($C^N_t$, eq. \ref{eq:seat_creation_new}). Destroyed seats comprise contractions of boards of permanent journals boards ($D^-_t$, eq. \ref{eq:seat_destruction_contracting}) and discontinued journals ($D^X_t$, eq. \ref{eq:seat_destruction_rate_exit}). The row ``0619'' reports values computed over the 13-year interval 2006-2019.}
\centering
\begin{tabular}[t]{crrrrrr}
\toprule
Year & $C^+_t$ & $C^N_t$ & $C_t$ & $D^-_t$ & $D^X_t$ & $D_t$\\
\midrule
1866 & 0 & 82 & 82 & 0 & 0 & 0\\
1876 & 26 & 25 & 51 & 4 & 0 & 4\\
1886 & 7 & 13 & 20 & 21 & 1 & 22\\
1896 & 33 & 63 & 96 & 29 & 0 & 29\\
1906 & 41 & 57 & 98 & 8 & 1 & 9\\
\addlinespace
1916 & 118 & 70 & 188 & 53 & 0 & 53\\
1926 & 118 & 212 & 330 & 100 & 4 & 104\\
1936 & 167 & 325 & 492 & 120 & 9 & 129\\
1946 & 217 & 329 & 546 & 223 & 134 & 357\\
1956 & 718 & 1,068 & 1,786 & 134 & 23 & 157\\
\addlinespace
1966 & 946 & 1,429 & 2,375 & 802 & 17 & 819\\
1976 & 1,944 & 2,848 & 4,792 & 822 & 36 & 858\\
1986 & 3,383 & 4,351 & 7,734 & 1,295 & 213 & 1,508\\
1996 & 4,698 & 7,035 & 1,1733 & 2,548 & 590 & 3,138\\
2006 & 6,965 & 11,835 & 18,800 & 4,074 & 1,057 & 5,131\\
\addlinespace
2012 & 10,325 & 8,865 & 19,190 & 4,137 & 475 & 4,612\\
2019 & 12,520 & 4,530 & 17,050 & 5,260 & 2,559 & 7,819\\
\addlinespace
0619 &16,152&14,144&30,306&4,315&2,182&6,497\\
\bottomrule
\end{tabular}

\label{tab:seat_creation_and_destruction_leve}
\end{table}

During the initial 80-year period (1866–1946), the total number of seats and members grew gradually. This expansion occurred alongside a stable median board size, indicating that growth was driven primarily by the entry of new journals that replicated the editorial board structures of existing ones. Throughout this period, most members held only a single editorial position, as evidenced by the low and stable seats-per-member ratio, which peaked at just 1.15 in 1936.

A decisive inflection point arrived between 1946 and 1956.  In this single decade, the total number of seats more than doubled (from 1,196 to 2,825), as did the community of members (from 1,074 to 2,484). However, this growth did not alter the fundamental structure of journal governance. Editorial boards remained limited in size and the practice of scholars holding multiple positions on boards of different journals, was not yet widespread.

The year 1956 seems to mark the end of this small-scale era. The following decades saw the consolidation of a new model of journal governance, characterized by larger editorial boards: after 1966, the median board size began a steady climb from 10 to 31 by 2019. The concurrent rise in the seats-per-journal indicator suggests the emergence of journals with significantly larger editorial boards. After 1976, the seats-per-member ratio increased from a historically stable 1.1 to reach 1.4, signaling the emergence of systematic interlocking editorship.

\begin{table}[htbp]
\centering
\caption{The rate of seat creation ($\dot{C}_t$, defined in eq. \ref{eq:gross_seat_creation}) is the sum of the rate of seat creation in journals expanding their boards ($\dot{C}^+_t$, eq. \ref{eq:seat_creation_expanding}) and the rate of seat creation generated by new journals ($\dot{C}^N_t$, eq.\ref{eq:seat_creation_rate_expanding}). The rate of seat destruction ($\dot{D}_t$, eq. \ref{eq:gross_seat_destruction}) is the sum of the rate of seat destruction in journals contracting their boards ($\dot{D}^-_t$, eq. \ref{eq:seat_destruction_rate_contracting}) and the rate of seat destruction due to discontinued journals ($\dot{D}^X_t$, eq. \ref{eq:seat_destruction_rate_exit}). The rate of seat growth ($\dot{S}_t$, eq. \ref{eq:seat_aggregate_growth_rate}) and the seat turnover rate ($\dot{T}^S_t$, eq. \ref{eq:seat_reallocation_rate}) are also shown. 
The row ``0619'' reports values computed over the 13-year interval 2006-2019.} 
\label{tab:seat_rates}
\begin{tabular}{ccccccccc}
  \toprule
Year & $\dot{C}^\mathbb{+}_t$ & $\dot{C}^\mathbb{N}_t$ & $\dot{C}_t$ & $\dot{D}^-_t$ & $\dot{D}^\mathbb{X}_t$& $\dot{D}_t$ & $\dot{S}_t$ & $\dot{T}^S_t$ \\ 
  \midrule
1866 & 0.000 & 1.000 & 1.000 & 0.000 & 0.000 & 0.000 & 1.000 & 1.000 \\ 
  1876 & 0.123 & 0.118 & 0.241 & -0.019 & 0.000 & -0.019 & 0.222 & 0.260 \\ 
  1886 & 0.027 & 0.051 & 0.078 & -0.082 & -0.004 & -0.086 & -0.008 & 0.164 \\ 
  1896 & 0.103 & 0.196 & 0.299 & -0.090 & 0.000 & -0.090 & 0.209 & 0.389 \\ 
  1906 & 0.086 & 0.119 & 0.205 & -0.017 & -0.002 & -0.019 & 0.186 & 0.224 \\ 
  \addlinespace
  1916 & 0.168 & 0.100 & 0.268 & -0.076 & 0.000 & -0.076 & 0.192 & 0.344 \\ 
  1926 & 0.111 & 0.200 & 0.311 & -0.094 & -0.004 & -0.098 & 0.213 & 0.409 \\ 
  1936 & 0.101 & 0.197 & 0.298 & -0.073 & -0.005 & -0.078 & 0.220 & 0.376 \\ 
  1946 & 0.099 & 0.149 & 0.248 & -0.101 & -0.061 & -0.162 & 0.086 & 0.410 \\ 
  1956 & 0.179 & 0.266 & 0.445 & -0.033 & -0.006 & -0.039 & 0.406 & 0.484 \\ 
  \addlinespace
  1966 & 0.131 & 0.198 & 0.329 & -0.111 & -0.002 & -0.113 & 0.216 & 0.442 \\ 
  1976 & 0.153 & 0.224 & 0.377 & -0.065 & -0.003 & -0.068 & 0.309 & 0.445 \\ 
  1986 & 0.148 & 0.190 & 0.338 & -0.057 & -0.009 & -0.066 & 0.272 & 0.404 \\ 
  1996 & 0.125 & 0.187 & 0.312 & -0.068 & -0.016 & -0.084 & 0.228 & 0.396 \\ 
  2006 & 0.116 & 0.197 & 0.313 & -0.068 & -0.018 & -0.086 & 0.227 & 0.399 \\ 
  \addlinespace
  2012 & 0.117 & 0.101 & 0.218 & -0.047 & -0.005 & -0.052 & 0.166 & 0.270 \\ 
  2019 & 0.112 & 0.040 & 0.152 & -0.047 & -0.023 & -0.070 & 0.082 & 0.222 \\ 
  \addlinespace
  0619 & 0.166 & 0.145 & 0.311 &-0.044 & -0.022 & -0.067&0.244 & 0.378\\
   \bottomrule
\end{tabular}
\end{table}

\begin{figure}
  \centering

  \begin{subfigure}[t]{0.90\textwidth}
    \centering
    \includegraphics[width=\linewidth]{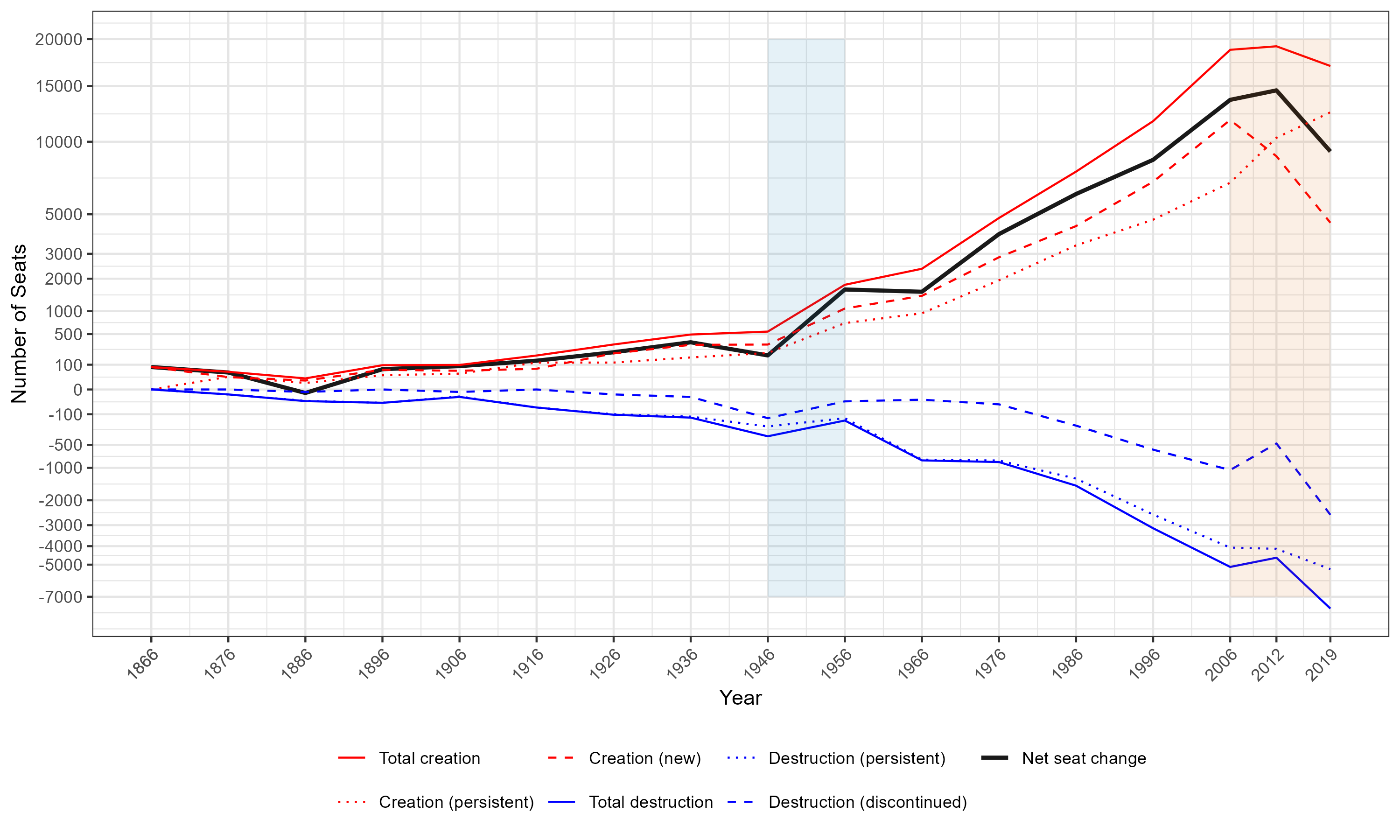}
    \caption{Seat creation, seat destruction and seat net change ($\Delta S_t$, eq. \ref{eq:seat_change}) in the period 1866-2019. See Table \ref{tab:seat_creation_and_destruction_leve} for the references to the analytical definitions.}
    \label{fig:panelA_seat_dynamics}
  \end{subfigure}
  \hfill
  \begin{subfigure}[t]{0.90\textwidth}
    \centering
    \includegraphics[width=\linewidth]{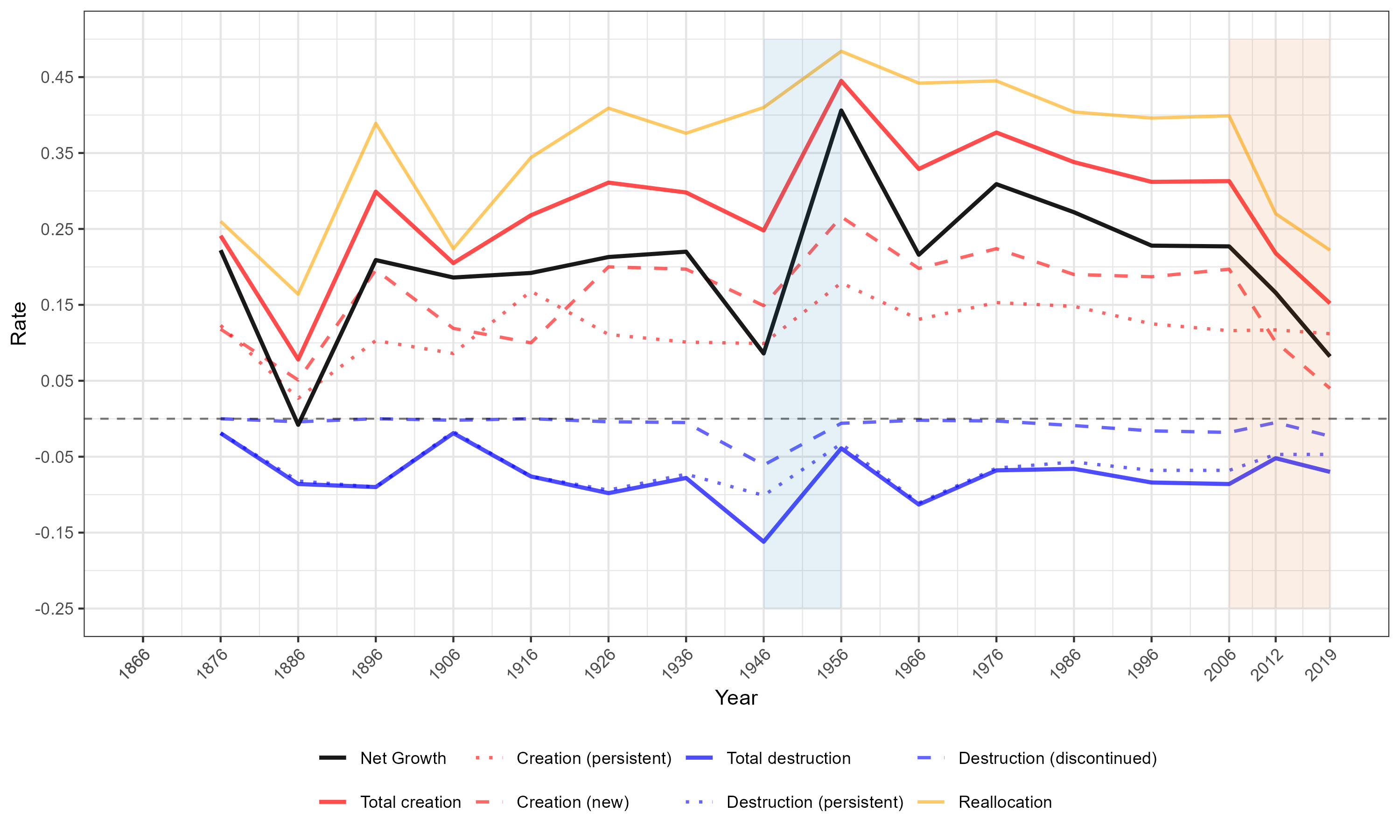}
    \caption{Normalized rates of seat net growth, creation, destruction, and turnover in economics journals (1866-2019). See Table \ref{tab:seat_rates} for the references to the analytical definitions.}
    \label{fig:panelB_seat_}
  \end{subfigure}

  \caption{Seat creation and destruction (1866-2019).}
  \label{fig:seats_flows}
  
\end{figure}

The longitudinal data on flows allow for a more precise periodization. Table \ref{tab:seat_creation_and_destruction_leve} reports the absolute values of the flows, Table \ref{tab:seat_rates} and Figure \ref{fig:seats_flows} show the rates of seat growth, creation, destruction, and reallocation.

The first period (1866-1946) is characterized by a volatile growth, with peaks in the rate of seat creation, interspersed with phases of stagnation, such as in 1886 where creation and destruction nearly offset. The steady expansion started in 1896 was stopped in 1946, which records the peak in the destruction rate \((\dot{D}_t = 0.162)\).

The decade ending in 1956 marks the absolute peak of the series for the rates of net seat growth, seat creation, and seat turn-over. This year inaugurated the 50-year expansionary phase that followed.

From 1956 until 2006, the editorial boards of economics journals were characterized by persistent expansion, driven by positive but decreasing rates of net seat creation. This growth was primarily fueled by the creation of new seats through the entry of new journals, which continuously exceeded the creation of seats due to the expansions of  editorial boards of existing journals. Throughout this period, the seat destruction rate due to journal exits remained persistently low, while seat turnover continuously decreased. This phase ended in 2006.

The most recent period appears to mark a structural break. When the 13-years between 2006 and 2019 are observed as a whole, the rate of seat growth appears lower than in large part of the previous expansionary decades. A clearer picture emerges when the period is split into two subperiods 2006-2012 and 2012-2019. The seat growth rate fell to its secular minimum in both intervals. A key change in the composition of net growth also emerges: the main driver of seat expansion changed. Unlike the post-WWII expansionary phases, growth is now due primarily to the enlargement of existing journals' editorial boards, rather than to seats generated by new journals. At the same time, the seat reallocation rate also reached minimum values. In the seven years between 2012 and 2019, the net growth rate reached the lowest value in the entire series except for 1886, indicating a severe contraction of the system's dynamism. 

\begin{figure}[htbp]
    \centering
    \includegraphics[width=0.9\textwidth, height=0.6\textheight, keepaspectratio]{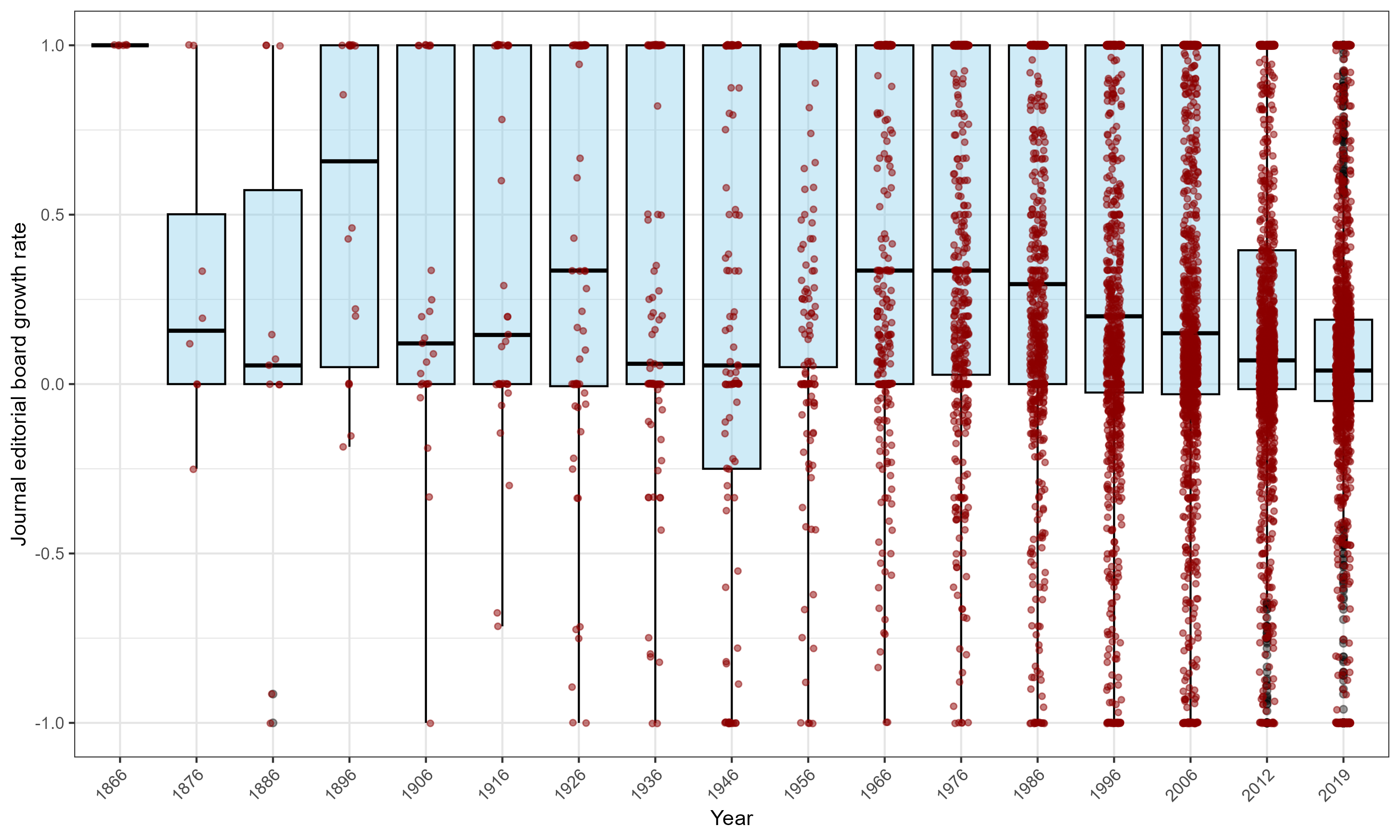}
    \caption{Distribution of journals according to their board growth rate ($\Delta s_{j,t}$, as defined in eq. \ref{eq:journal_board_growth}) in the period 1866-2019.}
    \label{fig:journal_seat_growth_boxplots}
\end{figure}

Figure \ref{fig:journal_seat_growth_boxplots} breaks down the aggregate statistics by showing the distribution of journal-level seat growth rates ($\dot{g}_{j,t}$) over time. The boxplot confirms the periodization observed at the aggregate level. The pre-1946 period shows pronounced volatility, with substantial positive skewness and numerous outliers. This pattern reflects the formative phase of the field, where journal creation and expansion were frequent, while exit and contraction were relatively rare.

The 1946 distribution displays marked negative skewness, capturing the unique contraction following World War II. 

The 1956 distribution emerges with dramatic positive skewness due to extensive journal entry, to the point where the median value of the seat growth rate calculated at a journal level reached its theoretical maximum of 1. During the subsequent 1956-2006 expansionary era, distributions maintained consistent right-skewness, though with reduced extreme values of entries of new journals compared to the immediate post-war year. The stability of the interquartile ranges throughout this period indicates persistent heterogeneity in the fortunes of journals.

The contemporary period (2012-2019) shows a notable shift toward distributional symmetry, with tighter clustering around the median and a strong diminution of journal entries. This compression suggests that the field is transitioning toward a new equilibrium with constrained growth opportunities and balanced expansion-contraction dynamics.

\subsection{The dynamics of membership on editorial boards}

\begin{figure}[htbp]
    \centering

    \begin{subfigure}[t]{0.9\textwidth}
     \includegraphics[width=\linewidth]{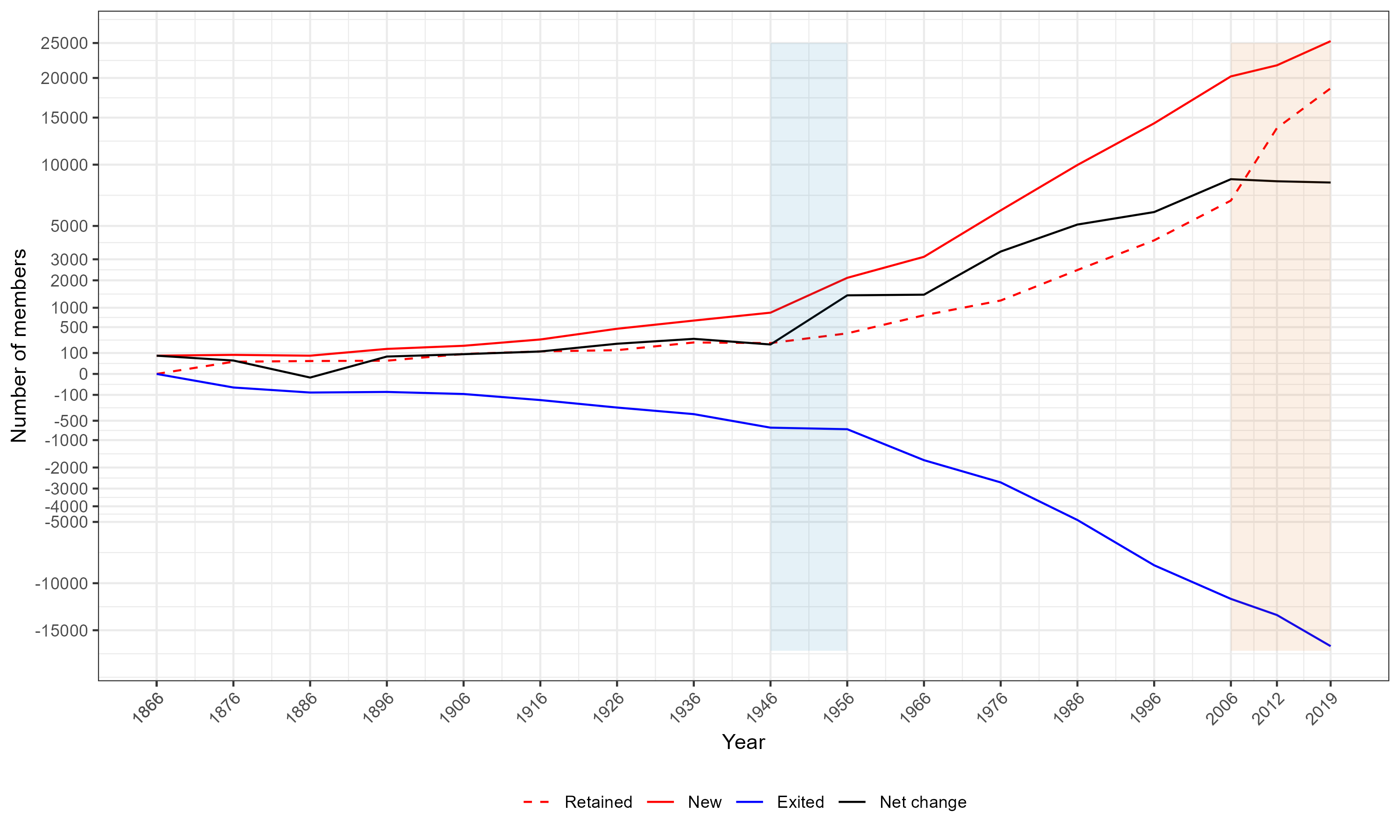}
     \caption{Global members dynamics by year (1866-2019).}
    \label{fig:panelamember_dynamics}
    \end{subfigure}
    \hfill

    \begin{subfigure}[t]{0.9\textwidth}
     \includegraphics[width=\linewidth]{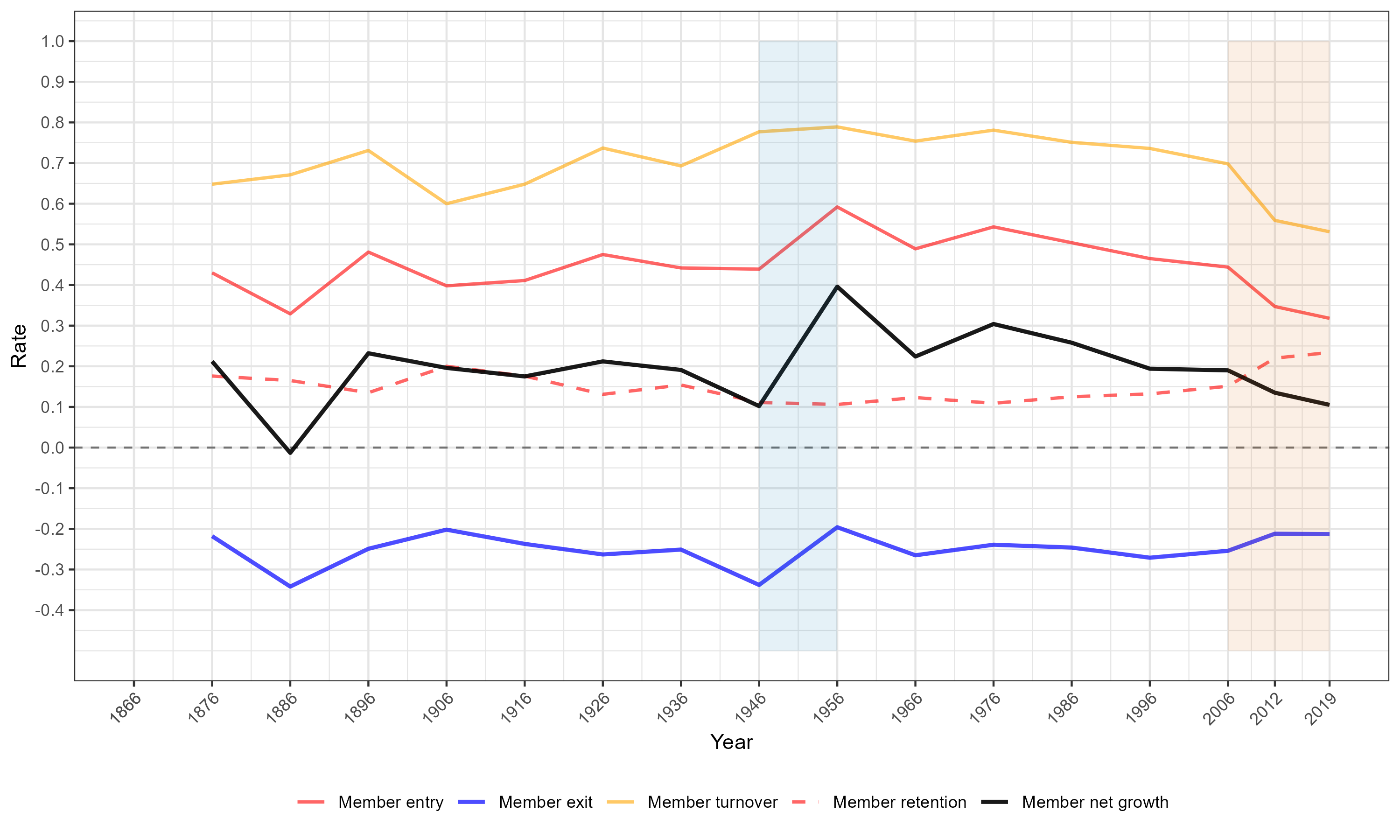}
     \caption{Normalized rates of member entry ($\dot{M}^N_t$), member exit ($\dot{M}^X_t$), member retention ($\dot{M}^R_t$), member net growth ($\dot{M}_t$), and member turnover ($\dot{T}^M_t$). See Table \ref{tab:member_dynamics} for the references to the analytical definitions.}
    \label{fig:panelBmember_dynamics}
    \end{subfigure}
  \caption{Member creation and member exit (1866-2019).}
  \label{fig:members_aggregate}
   
\end{figure}

Table \ref{tab:member_dynamics} and Figure \ref{fig:members_aggregate} present the evolution of the membership dynamics on editorial boards. The data reveal a remarkable transformation from the nascent phase of economics as an autonomous discipline to the contemporary consolidated structure.

The initial phase (1866-1906) exhibits volatility for all indicators. The relatively high entry rates offset by exit rates resulted in moderate net growth. In absolute terms, the community of editorial board members grew from just 76 members in 1866 to 272 members by 1906 (see Table \ref{tab:seats_members_demography}). The period (1916-1946) maintained similar patterns, though with gradually expanding absolute numbers of members, reaching 1,074 members by 1946. 

The year 1956 marks a structural break also for membership, culminating in the highest entry and turnover rates, the lowest retention rate, and the highest exit rate in absolute value for the entire series. 

After this peak, the field embarked on a sustained expansionary phase. During the subsequent decades, up to 2006, the discipline witnessed a tenfold expansion in the number of scholars serving as editors, which increased from 2,484 in 1956 to 27,075 in 2006. This expansion appears to be the result of a long-term steady growth phase, characterized by positive but gradually declining member entry rates, slowly increasing member retention rates, and slowly declining member turnover rates. This dynamics is consistent with a process of consolidation and stabilization of the communities of economists around journals.

The most recent period marks a change in the dynamics of membership as well. When observed as a whole, the 13-year span from 2006 to 2019 shows a member net growth rate of 0.237, which is higher than those of the two preceding decades. As anticipated, membership dynamics are the most sensitive to the length of the observation interval. Indeed, when the sub-periods 2006–2012 and 2012–2019 are examined separately, both expanded at the lowest rates recorded in the post-WWII era. Similarly, when observing the member turnover rate over the 2006-2019 interval, the figure ($0.713$) is only slightly lower than in previous decades. When observing the two sub-periods separately, a decline in the member turnover rate becomes apparent, even reaching the lowest value ($0.531$) of the entire series in the 2012-2019 decade. 

Despite the extended timeframe, the member retention rate between 2006 and 2019 remains the second highest in the post-WWII period, even if it is slowly declining compared to the previous decade. However, when the two subperiods are considered separately, the retention rates are the highest of the entire series. 
The member exit rates observed throughout the contemporary period are the smallest in absolute value since the expansion years that began in 1956, regardless of the time interval considered.

The combination of these patterns over the most recent years suggests a shift in the field of economics toward an editorial system characterized by journals governed by more stable and more closed editorial communities.

\begin{table}

\caption{Global members dynamics by year. Member counts: retained ($M^R_t$), new ($M^N_t$), exited ($M^X_t$); membership rates: retention ($\dot{M}^R_t$, as defined in eq. \ref{eq:member_retention_rate}), entry ($\dot{M}^N_t$, eq. \ref{eq:member_entry_rate}), exit ($\dot{M}^X_t$, eq. \ref{eq:member_exit_rate}), net growth ($\dot{M}_t$, eq: \ref{eq:member_growth_rate}), and turnover ($\dot{T}^M_t$, eq. \ref{eq:member_turnover_rate}); $\rho_t$ is the share of gross seat creation covered by new members (EQ. \ref{eq:new_member_coverage}). The row ``0619'' reports values computed over the 13-year interval 2006-2019. All rates are normalized.}
\label{tab:member_dynamics}
\centering
\begin{tabular}[t]{rrrrrrrrrr}
\toprule
Year & \multicolumn{1}{c}{$M^R_t$} & \multicolumn{1}{c}{$M^N_t$} & \multicolumn{1}{c}{$M^X_t$} & \multicolumn{1}{c}{$\dot{M}^R_t$} & \multicolumn{1}{c}{$\dot{M}^N_t$} & \multicolumn{1}{c}{$\dot{M}^X_t$} & \multicolumn{1}{c}{$\dot{M}_t$} & \multicolumn{1}{c}{$\dot{T}^M_t$} & \multicolumn{1}{c}{$\rho_t$}  \\
\midrule
1866 & 0 & 76 & 0 & 0.000 & 1.000 & 0.000 & 1.000 & 1.000& 0.927\\
1876 & 34 & 83 & 42 & 0.176 & 0.430 & -0.218 & 0.212 & 0.648 & 1.627\\
1886 & 38 & 76 & 79 & 0.165 & 0.329 & -0.342 & -0.013 & 0.671 & 3.800\\
1896 & 40 & 143 & 74 & 0.135 & 0.481 & -0.249 & 0.232 & 0.731 & 1.490\\
1906 & 91 & 181 & 92 & 0.200 & 0.398 & -0.202 & 0.196 & 0.600& 1.847\\
\addlinespace
1916 & 116 & 271 & 156 & 0.176 & 0.411 & -0.237 & 0.175 & 0.648 & 1.441\\
1926 & 129 & 466 & 258 & 0.131 & 0.475 & -0.263 & 0.212 & 0.737 & 1.412\\
1936 & 226 & 650 & 369 & 0.154 & 0.442 & -0.251 & 0.191 & 0.693& 1.321\\
1946 & 217 & 857 & 659 & 0.111 & 0.439 & -0.338 & 0.102 & 0.777 & 1.570\\
1956 & 376 & 2,108 & 698 & 0.106 & 0.592 & -0.196 & 0.396 & 0.789 & 1.180\\
\addlinespace
1966 & 786 & 3,132 & 1,698 & 0.123 & 0.489 & -0.265 & 0.224 & 0.754 & 1.319\\
1976 & 1,231 & 6,107 & 2,687 & 0.109 & 0.543 & -0.239 & 0.304 & 0.781 & 1.274\\
1986 & 2,465 & 9,968 & 4,873 & 0.125 & 0.504 & -0.246 & 0.258 & 0.751 & 1.289\\
1996 & 4,074 & 14,342 & 8,359 & 0.132 & 0.465 & -0.271 & 0.194 & 0.736 & 1.222\\
2006 & 6,862 & 20,213 & 11,554 & 0.151 & 0.444 & -0.254 & 0.190 & 0.698 & 1.075\\
\addlinespace
2012 & 13,804 & 21,746 & 13,271 & 0.220 & 0.347 & -0.212 & 0.135 & 0.559 & 1.133\\
2019 & 18,629 & 25,286 & 16,921 & 0.234 & 0.318 & -0.213 & 0.105 & 0.531 & 1.483\\
\addlinespace
0619 & 10,179  & 33,736 & 16,896 & 0.143 & 0.475 & -0.238 & 0.237 & 0.713 & 2.385\\

\bottomrule
\end{tabular}
\end{table}

\begin{sidewaysfigure}
  \centering
    \includegraphics[width=\textwidth, height=0.7\textheight, keepaspectratio=false]{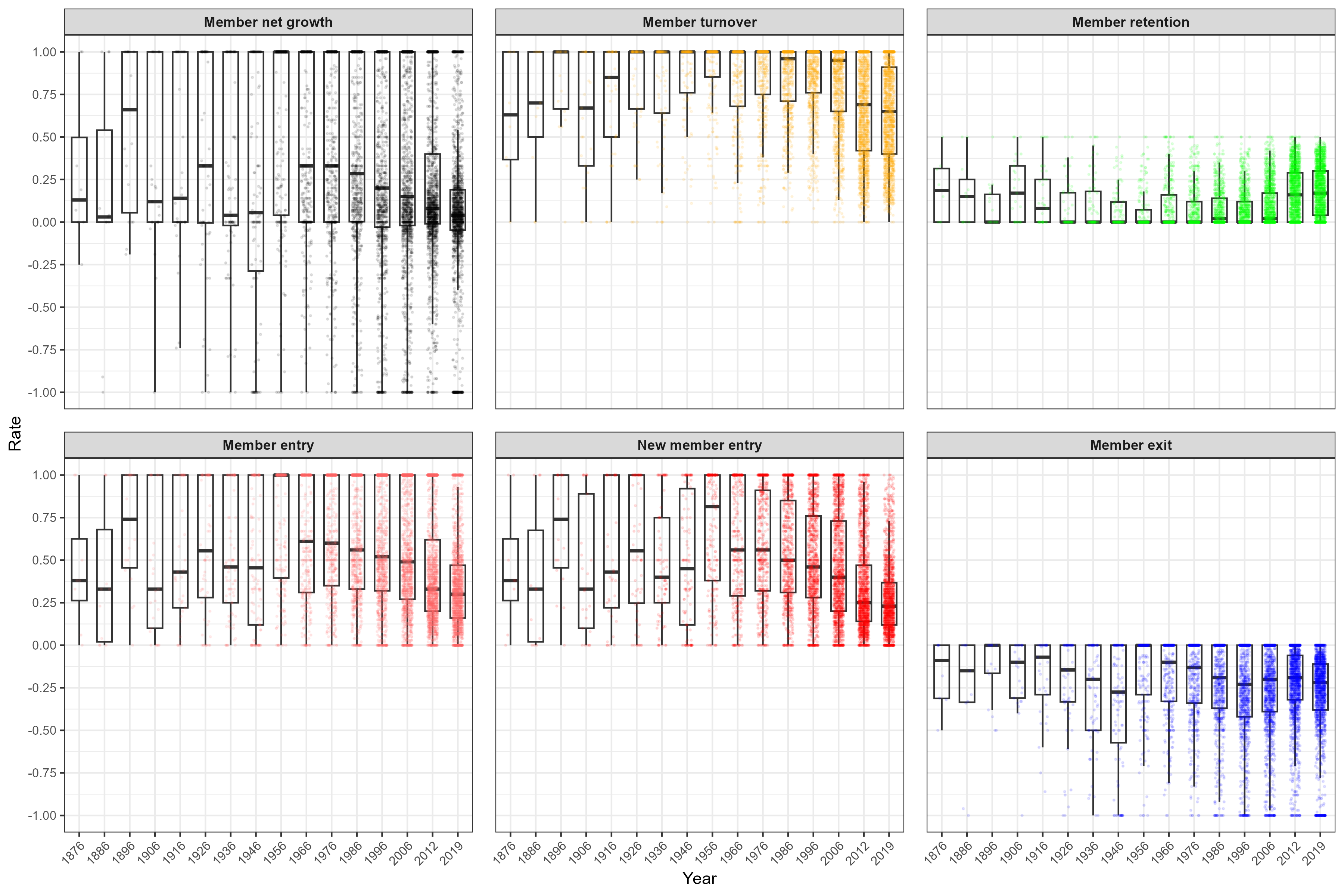}
    \caption{Distribution of journal-level member dynamics (1866--2019). 
The six panels display boxplots (with jittered points) showing the distribution of journals 
according to the following normalized rates: member net growth ($\dot{m}_{j,t}$, as defined in eq. \ref{eq:member_growth_journal}), 
member turnover ($\dot{T}^m_{j,t}$, eq. \ref{eq:turnover_member_rate_journal}), member retention ($\dot{m}^R_{j,t}$, eq. \ref{eq:member_retention_rate_journal}), 
member entry ($\dot{m}^n_{j,t}$, eq. \ref{eq:new_member_rate_journal}), new member entry ($\dot{m}^N_{j,t}$, eq. \ref{eq:New_member_rate_journal}), 
and member exit ($\dot{m}^x_{j,t}$, eq. \ref{eq:exit_member_rate_journal}). }
    \label{fig:member_journal_level_rates_distribution}
\end{sidewaysfigure}

As discussed previously, for members of the editorial boards, the aggregate data and rates are not merely the sum of data and rates calculated at journal level. The six panels of Figure \ref{fig:member_journal_level_rates_distribution} summarize the journal-level analysis of member flows. Each panel displays the annual box-plot distribution of journals for one of the six indicators considered. Collectively, these indicators confirm and reinforce at the journal level the periodization previously observed at the aggregate level. 

The first indicator, the net growth rate of the members ($\dot{m}_{j,t}$, defined in eq. \ref{eq:member_growth_journal}), confirms the volatility of the period 1866–1946, while also revealing a consistent right skewness in the distributions of journals. The year 1946 stands out with a pronounced left skew, likely reflecting contraction or disruption in editorial boards in the decade of WWII. 
As suggested by the aggregate analysis, 1956 emerges as a clear turning point, characterized by extreme right skewness and a median value reaching the theoretical maximum for net growth. The subsequent decades (1966–2006) are marked by persistently right-skewed distributions, coupled with a progressive decline in median net growth. 
Finally, the two most recent periods show a notable shift toward distributional symmetry, accompanied by a continued decrease in median growth, which reached its lowest point (0.04) in 2019. 

The distribution of journals according to the second indicator, the turnover rate of members ($\dot{T}^m_{j,t}$, defined in eq. \ref{eq:turnover_member_rate_journal}), shows a progressive reduction in the inter-quartile range from the beginning to 1956, accompanied by consistently extreme right skewness with median values at or very near to the theoretical maximum. Until 2006, this low variability persisted; however, since 2006, variability began to reemerge. The two most recent sub-periods are then characterized by distinctly higher dispersion and lower median turnover values. The aggregate tendency toward reduced member change is thus clearly confirmed at the journal level.

The distribution of journals according to the third indicator, the retention rates of members ($\dot{m}^R_{j,t}$, defined in eq. \ref{eq:member_retention_rate_journal}), which is defined in the range $[0,0.5]$, shows marked variability until 1916. From that point onward, variability declined and the distribution compressed toward zero. By 1956 the median had reached its theoretical minimum, and the inter-quartile range had narrowed to its lowest point. Subsequently, the variability exhibited a gradual increase, and in the two most recent periods the member retention rate reached its highest historical levels. 

The distribution of the journals according to the fourth indicator, the member entry rate ($\dot{m}^n_{j,t}$, defined in eq. \ref{eq:new_member_rate_journal}), confirms the initial period of instability characterized by high variability and right skewness. The year 1956 again emerges as a turning point, marked by extreme right skew and a median value at its theoretical maximum. The following decades show a steady decline in median entry rates, with the overall minimum reached in 2019. Finally, the distributions for 2012 and 2019 are characterized by strong compression of skewness and symmetry around these notably lower median values.

The fifth indicator is the new member entry rate ($\dot{m}^N_{j,t}$, defined in eq. \ref{eq:New_member_rate_journal}) which considers only scholars who join the editorial board of a journal and are completely anew for the editorial system of the observed year. The distribution of journals according to this indicator reinforces the pattern observed for the overall member entry rate. This includes a volatile initial period lasting several decades; a distinct peak in 1956, marked by the highest median value and pronounced right skewness; and a subsequent phase of slowing entry rates with reduced variability. The years 2012 and 2019 exhibited the lowest median values for new-member entry, with the distribution becoming more symmetric and more tightly concentrated around these lower medians.

Finally, the distribution of journals according to the sixth indicator, the member exit rates ($\dot{m}^x_{j,t}$, defined in eq. \ref{eq:exit_member_rate_journal}), defined in the range $[-1,0]$, also confirms the volatility of the first decades. The renewal of 1956 was accompanied by the highest  median rate of member exit which equaled the theoretical maximum. This finding indicates that the peak turnover rate of the members in that year was entirely driven by the member entry rate. In the subsequent decades up to 2006, a gradual and steady increase in the absolute values of the exit rate was observed. In contrast, the two most recent periods exhibit an inversion of this trend by a decrease in both the median member exit rate and its interquartile range.

\section {Discussion}

Scholarly journals are institutionalized spaces in which the social and intellectual relations that structure a discipline can be observed. The editorial board is the main organizational structure of a journal and, by and large, represents its governance body. However, this governance structure varies considerably across journals and over time. For instance, there are journals with a quasi-monocratic structure consisting of a single member of the editorial board, and journals with boards comprising hundreds of members. Similarly, internal governance rules, including those for functioning and decision-making, are highly variable both across journals and historically.

Anecdotal evidence suggests that, before the generalized diffusion of peer review, editorial boards often decided directly about the publication of articles \citep{Baldwin}. This decision could be made by the editor-in-chief, the entire board, or a specialized editor. The introduction of peer review further complicates the decision-making process by increasing the differentiation among the roles of board members. At one extreme, some boards leave the final decision on publication to the editor-in-chief, who also selects referees directly. At the other end of the spectrum, decision-making is completely decentralized to a board member, such as an associate editor, who also selects the referees. Between these poles lies a wide spectrum of intermediate configurations . These include cases where an associate editor selects referees, but the final decision rests with the editor-in-chief, or cases where the process is managed collectively by a relatively small, collegial board. This variety of roles and functions means that the title of a member of editorial boards, such as ``editor'', ``associate editor'', or ``member'', often entails different responsibilities and powers across journals and historical periods. This variety makes it extremely difficult to analyze the role of the board in editorial decisions (see e.g. \citet{Horbach, Kochetkov}).

In order to analyze journal governance structures and their evolution, we propose adopting a structural and quantitative longitudinal perspective, using a stock-flow approach. This approach is largely inspired by labour market stock-flow analysis, but incorporates several key conceptual and operational differences. While labor market dynamics examine the demographics of firms and the stock-flow of jobs, our analysis of journals operates across three layers: the demographics of journals, the dynamics of editorial seats, and the dynamics of membership. The first two layers are analytically straightforward and can be examined consistently when the analysis is conducted at the journal level and when it is conducted at aggregate levels, such as the set of journals representative of a discipline. However, the layer of member dynamics requires an analytical shift when moving from the journal level to the aggregate level. Indeed, at the journal level, the notions of a ``new member'' and an ``exiting'' member become decoupled from their aggregate definitions: a member new to a specific journal may not be new for the entire set of journals, and an exit from one journal may represent a move to another rather than leaving the system.

In the first part of the paper, we have developed the first systematic theoretical stock-flow framework for analyzing editorial boards, defining the relevant indicators for data observed at fixed-length periods. We therefore propose a strategy for normalising indicators to adapt them to the processing of data referring to periods of variable duration, as in our empirical case.

This framework is applied to the analysis of the editorial boards of economics journals over the very long run, covering the evolution of the discipline from the foundation of its first journals in 1866, through its development as an autonomous field, to its current configuration. The data analyzed enable for the first comprehensive depiction of the evolution of journal publishing and its governance for an entire scholarly discipline over such an extended period.

The GOELD database used here was constructed from the list of economics journals maintained by the American Economic Association. This list, which included approximately 1,700 journals in 2019, comprises publications indexed in the \textit{EconLit} database, currently the most comprehensive bibliographic database for economics. It is therefore far more inclusive than the limited lists of economics journals indexed in the \textit{Web of Science} or \textit{Scopu}s. However, it also includes many journals that general bibliographic databases might more properly classify as belonging to other fields, such as statistics. The main advantage of using this list is that it allows us to adopt the perspective of the largest academic society in the field when defining the boundaries of what constitutes a journal of interest to economists. This perspective is, of course, not exhaustive. It likely underrepresents journals published in languages other than English or from the Global South, as well as journals positioned at the intersection of fields that critically engage with economics. For instance, the list of heterodox economics journals, maintained by Jakob Kapeller and Niklas Klann,\footnote{ https://web.archive.org/web/20251203143852/https://heterodoxnews.com/hed/journals.html.} includes many titles not covered by \textit{EconLit}. In summary, our analysis adopts the definition of ``journals of interest to economists'', or, in brief, the definition of  ``economics journal'' provided by the American Economic Association. While this may represent a limitation, it also provides a valuable analytical benchmark: the resulting list is not exhaustive, but is representative of what mainstream economists consider to be economics journals.

The most prominent feature of the data considered is the continuous expansion of the publishing landscape of economics from 1866 onward, measured by the number of journals, seats, and members. Three periods can be identified, with the decade following the end of the World War II representing the most significant point of discontinuity. 

The first period is from the beginning to 1946, when economics was established as an autonomous discipline, remaining small in scale. By 1936, fewer than 80 journals were active, with about a thousand editorial board seats, which fewer than nine hundred scholars occupied. While the decades from 1800 to 1936 were characterized by the volatility of seats and members, the decade encompassing World War II saw relatively high turnover in journals, seats, and members. However, the overall size of the publishing landscape remained substantially unchanged by 1946. 

The decade between 1946 and 1956 marked a clear discontinuity in the evolution of economics. These ten years saw the highest growth rates in the number of journals, seats, and scholars, accompanied by dramatic membership turnover and the most significant influx of new members into the economics publishing arena. This period likely represents the moment when economics shifted toward a big science model. The discipline became increasingly relevant to policy makers, and post-war research funding from Western institutions was also directed toward economics. This shift coincided with the relocation of the center of the discipline from the United Kingdom to the United States.  Concurrently, significant intellectual and institutional transformations took place including the consolidation of the neoclassical synthesis as the prevailing paradigm and the establishment of entities such as the International Monetary Fund. These development provided fertile ground for the sustained growth of economics publishing sector \citep{bender, Fourcade_2001, marchionatti2, Marchionatti3}. 

The subsequent decades, up to 2006, were marked by the uninterrupted expansion of the publishing landscape. Growth rates slowed compared to the 1956 peak; nevertheless, the number of journals, seats, and members continued to expand at a steady state. Until 2012, growth was predominantly driven by new journal creation, which consistently contributed more seats than the expansion of the size of the boards of existing journals. It should be noted that the growth rate of seats tended to exceed that of individual members, thereby indirectly signaling the rise of interlocking directorates. 

The contemporary period from 2006 to 2019 seems to mark a structural break. The net growth rate of journals decelerated to nearly zero in 2019. The growth in the number of seats was primarily driven by the expansion of editorial boards within established journals. This growth, however, has decelerated compared to the rates observed in previous decades. The entry of new members reached a historical minimum, as did the member turnover rate. The journal-level analysis of member flows lends further support to the evidence of a structural break. 
Member net growth and member turnover rates fell to unprecedented low levels, with median values approaching zero. Concurrently, the distributions of net growth, entry, and exit rates of members shifted toward greater symmetry and tighter concentration around these lower medians, indicating a homogenization of journal-level behaviors. The dynamics appears to be driven by two complementary forces: member retention rates climbed to their highest recorded levels, while entry rates for both new and existing members dropped to their lowest. The long-term gradual increase in exit rates reversed, confirming that the slowdown is primarily an entry-driven phenomenon.

In summary, since the 2008 global financial crisis, the economics publishing landscape has entered a phase marked by low flux and high membership stability. A similar slowdown and eventual halt in the growth of Scopus indexed journals during the 2010-2020 decade was documented by \citet{Thelwall_sud}. They attribute it to the rise and expansion of broad-scope mega-journals, such as \textit{PLOS ONE} and \textit{Scientific Reports}, which effectively saturated scholarly niches that might otherwise have been filled by new, specialized journals. This explanation, however, does not readily apply to economics. Indeed, economics has no specific mega-journals of its own, and its presence in interdisciplinary mega-journals remains marginal \citep{Siler_et_al_2019}. Consequently, the observed stagnation in economics publishing landscape demands an alternative, field-specific explanation.

This shift in the publishing landscape may constitute a transition toward ``bureaucratic growth'' and a concomitant decline in the organizational diversity and pluralism within the economics discipline. The determining factor behind this phenomenon can be traced back to a systemic intellectual defense reaction to criticisms leveled at the discipline in the aftermath of the 2008 financial crisis. The criticism is famously and exemplarily summarized in the question posed by Queen Elizabeth II to economists at the London School of Economics: ``Why did nobody see it coming?''.  The profession's response to the crisis was predominantly characterized as a technical challenge to be resolved within the discipline's established social and intellectual settings \citep{mirowski,aignerkapeller,Levy}. This response entailed an avoidance of a reconsideration of the discipline's foundational assumptions or institutional structures. Rather, it fostered a mechanism of epistemic closure and possibly insularity \citep{Truc_2023}.

Two peculiar elements in the organization of economics have contributed to the epistemic closure. The first is the role of elite general-interest journals, most notably the so-called ``Top Five'' (\textit{American Economic Review}, \textit{Econometrica}, \textit{Journal of Political Economy}, \textit{Quarterly Journal of Economics}, \textit{Review of Economic Studies}). These journals and their editorial boards function as the supreme gatekeepers of economic science. They define the methodological and intellectual standards for the entire field, confer ultimate academic legitimacy, and control access not only to the most prestigious publication slots but also to the best academic positions \citep{heckman2020}. Their editorial boards are the central nodes in the gatekeeper network that defines hierarchy, consensus, and control in economics \citep{baccini_re, Fourcade}. The second element is that economists are obsessed with rankings, especially of journals \citep{Mogstad, hylmo}. Many papers published in economics journals have developed rankings of economics journals (the most recent being \citet{ham}). Rather paradoxically, heterodox economists have also developed their own rankings \citep{Lee_2010, Cronin_2020}. While in many other fields citations are the main requirement for climbing the academic ladder, in economics, scholars predominantly seek to publish in the top five journals \citep{heckman2020, rossier}. Individual institutions and even entire countries use journal lists for recruitment and career advancement. Consequently, sustaining or enhancing the standing of an economic journal within the hierarchy of academic journals is a pivotal challenge for its editorial board. The pursuit of this objective can be achieved by strategically attracting distinguished scholars to serve on the editorial boards. Consequently, the bureaucratic growth that has been observed can be interpreted as the result of a competitive effort to consolidate existing resources and ascend a hierarchy that is dominated by the top five journals of economics \citep{baccini_re}. 

This pattern of intellectual and social closure finds its organizational expression in the long-term dynamics of editorial boards. Our analysis reveals a system that has decisively turned inward, consolidating incumbent journals and their gatekeepers while drastically curtailing the entry of new journals and scholars. The mechanism behind this shift is a reversal in the very engine of editorial growth: expansion is now driven by the enlargement of existing boards, rather than by the entry of new actors as was historically the case. The field has moved from a dynamic era of high-variance, entry-driven growth to a stable period of incumbent consolidation.

Such a narrowing of developmental trajectories gives rise to critical inquiries about the discipline's long-term intellectual vitality. All the more so as the incumbent elite shows little to no willingness to engage, in academic debates and in public discourse, with heterodox or disruptive ideas, that are frequently disregarded at best as misguided, at worst as pseudo-science. This systematic de-legitimization is reinforced by the established authority of journal gatekeepers. Their boundary work polices the frontiers of economics by defining what constitutes legitimate knowledge and policies, and who is entitled to produce it \citep{baccini_shape_2025}.

\section{Conclusion}

Despite their centrality in the scholarly landscape, we still know relatively little about the functioning of scientific journals and their contribution to the advancement of science. Existing literature has contributed to this knowledge by focusing on the composition of editorial boards, their interrelations, the study of scholarly communities gathered around journals, and, finally, their intellectual contents. Some studies have also explored the markets for journals. However, these strands of literature remain largely disconnected, and a comprehensive synthesis of the role of journals, both in contemporary science and over the long term, is still lacking.

In this perspective, the analysis of the internal governance of scholarly journals is central to understanding their gatekeeping function. Such an analysis requires a robust complementarity between quantitative and qualitative approaches. The last thing we need is more metrics without theory. The label of ``editormetrics'' proposed by \citet{mendonca} to define quantitative studies of editorial boards is sometimes used, but it should be avoided for two main reasons. The first is a matter of expediency: it is best to avoid hypes in the very definition of disciplines. The simple possibility of developing indicators based on a specific type of information is not sufficient to define a new specific subfield of research. In economics and sociology, where the study of board of directors of firms was largely developed, there is no subfield called ``directormetrics'' or anything similar. The second is more critical. The focus on ``metrics'' risks to narrow the investigative perspective about journal governance to the construction of new \textit{ad-hoc} indicators, which can be added to the plethora of evaluative indicators already collected over decades of scientometrics research. The focus should not be on building indicators \textit{per se}, but on developing clear research questions about the role of journals in science, their organization, the social communities surrounding them, and their intellectual contents.

This paper contributes to understanding journals as social institutions of science by systematizing the analysis of their governance structures in the long run through a stock-flow framework. To this end, the theoretical structure of the stock-flow approach is developed by highlighting the relationships between journal demographics, the stock-flows of position available in their boards, and the stock-flows of scholars who occupy these positions. The proposed analytical model has proved capable of identifying major changes in the long-term organization of the publishing landscape in economics, consistent with existing knowledge about the history of economic thought and its periodization. 

More broadly, this framework opens new research questions for the science of science by shifting the analysis of journal governance from a static to a dynamic perspective. It enables us to explore, for instance, whether major changes in scholarly publishing, such as the introduction of systematic peer review or the transition from small to large publishers, were accompanied by transformations in editorial board structure and composition. As for board composition, the stock-flow approach allows us to dynamically analyze it by tracking new, retained, and outgoing members over time. This can be used to explore whether entry patterns reproduce or challenge existing gender, institutional, and cultural profiles of journals, and whether changes stem from new journals or from internal turnover. And again, our findings show that in recent decades, the expansion of editorial communities in economics has slowed, likely due to a defensive reaction to the post-2008 critiques of economics that emerged in the wake of the financial crisis. Whether this slowdown is unique to economics or reflects a broader structural shift in scholarly publishing remains an open question.

Finally, this type of analysis requires complex and costly databases. The scientometrics community should therefore consider undertaking long-term projects to create structured editorial archives that can support future research in this area.

\section*{Declarations}

\begin{itemize}
\item Funding: the research is funded by the Italian Ministry of University, PRIN project: How economics is changing: A multilayer network analysis of the recent evolution of economics journals, between specialization and self-referentiality (1980-2020),  2022SNTEFP, PI: Alberto Baccini.
\item Conflict of interest/Competing interests: The author has no competing interests to declare that are relevant to the content of this article.
\end{itemize}

\bibliography{references}




\end{document}